\documentclass[12pt,preprint]{aastex}
\title {Galaxy Luminosity Functions from Deep Spectroscopic Samples of Rich Clusters}
\author {Daniel Christlein \& Ann I. Zabludoff}
\affil {Steward Observatory, The University of Arizona}
\affil {933 N Cherry Ave, Tucson 85721 AZ}
\email {dchristlein@as.arizona.edu \\ azabludoff@as.arizona.edu}
\date {\today}
\begin {document}
\begin {titlepage}
\shortauthors{Christlein \& Zabludoff}
\shorttitle{Galaxy Luminosity Functions in Clusters} 
\begin {abstract}

Using a new spectroscopic sample and methods accounting for spectroscoic sampling fractions that vary in magnitude and surface brightness, we present R-band galaxy luminosity functions (GLFs) for six nearby galaxy clusters with redshifts $4000 < cz < 20000$ km/s and velocity dispersions $700 < \sigma < 1250$ km/s. In the case of the nearest cluster, Abell 1060, our sample extends to $M_{R}=-14$ (7 magnitudes below $M^{*}$), making this the deepest spectroscopic determination of the cluster GLF to date. Our methods also yield composite GLFs for cluster and field galaxies to $M_{R}=-17$ ($M^{*}+4)$, including the GLFs of subsamples of star forming and quiescent galaxies. The composite GLFs are consistent with Schechter functions ($M^{*}_{R}=-21.14^{+0.17}_{-0.17}$, $\alpha=-1.21^{+0.08}_{-0.07}$ for the clusters, $M^{*}_{R}=-21.15^{+0.16}_{-0.16}$, $\alpha=-1.28^{+0.12}_{-0.11}$ for the field). All six cluster samples are individually consistent with the composite GLF down to their respective absolute magnitude limits, but the GLF of the quiescent population in clusters is not universal. There are also significant variations in the GLF of quiescent galaxies between the field and clusters that can be described as a steepening of the faint end slope. The overall GLF in clusters is consistent with that of field galaxies, except for the most luminous tip, which is enhanced in clusters versus the field. The star formation properties of giant galaxies are more strongly correlated with the environment than those of fainter galaxies.

\end {abstract}
\maketitle
\end {titlepage}

\section{Introduction} 

The galaxy luminosity function (GLF) is one of the basic statistics of the properties of galaxies, and variations of the GLF as a function of environment provide important constraints on any attempt to model galaxy evolution. Despite recent progress on the determination of the field GLF \citep{sloanlf,2dflf}, less is known about the GLF in high density environments like rich clusters. In particular, there is controversy as to whether 1) the GLF in clusters is universal, 2) whether the GLF of clusters differs from the field, and, 3) which galaxy populations are most responsible for any differences between environments. These questions are unresolved for several reasons: 1) large volume- or magnitude-limited redshift surveys have far better statistics on galaxies in more common, less dense environments than clusters, 2) most past analyses have depended on statistical background subtraction, which is sensitive to cosmic variance \citep{valotto}, to constrain the faint end, 3) past work has often used B-band magnitudes, which are sensitive to dust and recent star formation, and 4) comparisons among clusters, or between cluters and the field, have been inhomogeneous, with data galaned from different datasets. In this paper, we aim to address these problems with the first deep, spectroscopic determination of the R-band GLF for six clusters of galaxies and their associated fields.

Past studies examining the universality of the GLF in clusters have yielded conflicting results \citep{trentham1998,smithetal1997,driver1998}. A definitive answer to this question requires not only a reliable determination of the mean GLF in clusters, but also strong constraints on the GLFs of individual clusters to at least 3-4 magnitudes below $M^{*}$ (the characteristic magnitude at which the exponential cutoff at the bright end of the Schechter function \citep{schechter1976} begins to dominate over the power law describing the faint end) in order to constrain their scatter around the mean. Semi-analytic models that attempt to reproduce the GLF in clusters \citep{springel01} are so far limited to simulating relatively small numbers of systems and are thus affected by cosmic variance. An estimate of the observational variance of the cluster GLF will be of great relevance to judging the quality of such models.

It is known from other statistics, such as the morphology-density relation \citep{dressler80}, that the properties of galaxies in clusters are different from those in the field. Are these differences reflected in the shape of the GLF? This question has not been resolved unambiguously either. \citet {zabmul00} find a consistency between poor group and rich cluster GLFs, and \citet{muriel} find group GLFs to be consistent with the field GLF. \citet{christlein2000} finds a systematic and continuous variation of the GLF faint end slope over a range of environments that mostly covers poor groups from the Las Campanas Redshift Survey \citep{shectman96}, but does not have sufficient data to extend the analysis to high-mass systems. To make a proper comparison from the highest to lowest density environments requires a field GLF that has been obtained from the same survey and processed using the same criteria (such as star/galaxy separation and surface brightness limits). Such an approach guarantees internal consistency and avoids the problems associated with comparing GLFs across different surveys (for example, see \citet{sloanlf} for a discussion of the effect of different surface brightness cutoffs on GLF determinations).

A related question is whether it is possible to identify particular populations of galaxies ({\it e.g.}, early or late types) that individually show variations with the environment. This test would be a potentially even stronger constraint on galaxy evolution models than the total GLFs alone. Furthermore, we would like to know whether any differences between the field and cluster GLFs arise solely from the morphology-density relation, from mixing differently populations with universal, type-specific GLFs \citep{bromley}, or whether the type-specific GLFs themselves depend on environment.

In the present paper, we measure cluster GLFs for six nearby (cz$<$20000 km/s) clusters and their surrounding fields. Our study is based on a spectroscopic sample that includes 300-500 galaxies per cluster, thus making statistical background subtraction unnecessary. The samples extend to $M_{R}\approx-18$ ($M^{*}+3$) for the highest-redshift cluster, A3266, and to $M_{R}\approx-14$ ($M^{*}+7$) in the case of the lowest-redshift cluster, A1060. For the purposes of determining composite cluster and field GLFs and comparing them, we impose magnitude limits that typically restrict our analysis to $M_{R}\leq-17$. Of 1860 spectroscopically confirmed cluster members, 1563 are within the magnitude limits of this study.

The clusters in our sample span a range of velocity dispersions (700 $<$ $\sigma$ $<$ 1250 km/s), providing a significant baseline for studies of any variations in the shapes of the GLF with cluster environment. A set of 703 galaxies confirmed non-members from the cluster fields allow for a self-consistent comparison between the field and cluster GLFs.

Our study provides complementary results to an independent study, also based on a spectroscopic sample, by \citet{depropris} of cluster GLFs in the $b_{J}$-band. Our R-band results are not strongly biased by the current or most recent star formation history of a galaxy or by its dust content. The R-band is more sensitive to the total stellar mass than bluer bands. At the same time, the choice of the R-band allows for deeper samplings of the GLF than studies in the infrared \citep{depropris99,kochanek}, which provide an even better (though not perfect) representation of the total stellar mass \citep{ericroelof}.

In \S 2, we first describe the six cluster samples. We then review how we obtained and reduced the data, including the procedures used to compile the detection catalog. We took particular care to optimize our photometry and star/galaxy separation. We describe the calculation of the cluster and field luminosity functions in \S4, with special care given to the treatment of the sampling fraction, $f_{s}$. We present our results in \S 5, including the six cluster GLFs, split into subsamples by their spectral properties, as well as the composite cluster and field GLFs. Finally, we test the six cluster samples for consistency with the composite cluster GLF to determine if the latter can serve as a common parent distribution for the galaxy populations in our clusters, and we compare the field and cluster composite GLFs using several tests. 

\section{The Data}

\subsection{The Sample}

Our dataset is a spectroscopic survey of galaxies in the fields of six low-redshift (cz $\leq$ 20,000 km/s) galaxy clusters. These clusters were selected based on 1) their visibility from Las Campanas, 2) the availability of some prior spectroscopic and X-ray data in the literature, 3) their redshifts, which allowed us to sample a large fraction of the virial radius with the 1.5$\times$1.5 degree field of the fiber spectrograph field, and 4) their range of velocity dispersions, which suggest a wide range of virial masses. The properties of these clusters (for $H_{0}=100$ km s$^{-1}$ Mpc$^{-1}$, $\Omega_{m}=0.3$ and $\Omega_{\Lambda}=0.7$, as applied throughout this paper) are given in Table 1. In this table, $\Delta m$ is the distance modulus that we adopt, $\sigma$ the internal velocity dispersion of the cluster, and $r_{sampling}$ the spectroscopic sampling radius. 

\begin{deluxetable}{lrrrrcrr}
\tabletypesize{\scriptsize}
\tablecaption{The Cluster Sample}
\tablewidth{0pt}
\tablehead{
\colhead{Cluster}&\colhead{N}&\colhead{$\bar{cz}$ [km/s]}&\colhead{$\Delta m$ [mag]}&\colhead{$cz$ range [km/s]}&\colhead{$\sigma$ [km/s]}&\colhead{$r_{sampling}$ [Mpc]}
}
\startdata
A1060&252&$3683\pm46$&32.85&2292  - 5723&$724\pm31$&0.48\\
A496 &241&$9910\pm48$&35.03&7731  - 11728&$728\pm36$&1.24\\
A1631&340&$13844\pm39$&35.78&12179 - 15909&$708\pm28$&1.71\\
A754 &415&$16369\pm47$&36.16&13362 - 18942&$953\pm40$&2.00\\
A85  &280&$16607\pm60$&36.19&13423 - 19737&$993\pm53$&2.03\\
A3266&331&$17857\pm69$&36.35&14129 - 21460&$1255\pm58$&2.19\\
\enddata
\end{deluxetable}

\subsection{Spectroscopy}

We selected targets for the spectroscopic sample by running the FOCAS software \citep{jarvistyson} on Digitized Sky Survey (DSS) plates of the survey region to detect diffuse objects and obtain approximate photometry in the $b_{j}$ band. We describe in \S 3.1 the effect of target selection in $b_{j}$ on the determination of R-band GLFs. We the prioritized these targets in order of increasing magnitude. We carried out the spectroscopic observations with the multifiber spectrograph \citet{shectman92} at the 2.5m DuPont telescope at the Las Campanas Observatory (LCO), targetting each field multiple times to ensure that no galaxies were lost to fiber crowding problems.

We extract, flat-field, wavelength-calibrate and sky-subtract (based on the flux normalization of the 5577 \AA\, 5890 \AA\, and 6300 \AA\ night sky lines) each spectrum. The spectra have a resolution of $\sim$ 5-6 \AA, a pixel scale of $\sim$ 3 \AA, and a wavelength range of 3500-6500 \AA. The average signal-to-noise ratio (S/N) in the continuum around the $H\beta$ $\lambda 4861$, $H\gamma$ $\lambda 4340$, and $H\delta$ $\lambda 4102$ absorption lines is typically $\sim 8$ (calculated by determining the ratio of the mean square deviation about the continuum at the absorption line, after excluding the absorption line and any nearby sky lines). The fiber aperture is $3.5^{\prime\prime}$.

We determine the radial velocities using the cross-correlation routine XCSAO and the emission-line finding routine EMSAO in the RVSAO package in IRAF \citep{minkwyatt}. The velocities in Table 2 are either emission-line velocities, absorption-line velocities, or a weighted average of the two (for a discussion of the cross-correlation templates and the spectral lines typically observed, see \citet{shectman96}, \S 2.2; \citet{lin95}). We compute velocity corrections to the heliocentric reference frame with the IRAF/HELIO program.

We estimate the velocity zero-point correction and external velocity error by comparing our velocities with H I velocities from NED. Fig. 1 shows the residuals for 61 galaxies (22 galaxies in A1060 from \citet{mcmahon} and 39 galaxies in the fields of poor groups, which were observed with the same instrumental setup \citep{zabmul98}) as a function of our internal velocity error estimate. We use only those H I velocities with quoted errors of $<$30 km s$^{-1}$. The mean residual of 11 km s$^{-1}$ ({\it solid line}) is small compared with the rms deviation of the residuals ($\sim$76 km s$^{-1}$) and is consistent with the mean residual of the 336 stars (52 km s$^{-1}$) that were serendipitously observed with the same instrument ({\it dashed line}). Therefore, we do not apply a zero-point correction to the velocities.

We adopt the rms deviation of the residuals ($\sim80$ km s$^{-1}$), which is constant over the range of internal errors, as the true velocity error when the internal or NED error is smaller than 80 km s$^{-1}$. Otherwise, we list the internal or NED error. Our error estimates are consistent with the average external error estimate of 70 km s$^{-1}$ for the Las Campanas Redshift Survey \citep{shectman96} and with \citet{zabmul98}, which both employ the same fiber spectrograph setup. 

\begin{figure}
\plotone{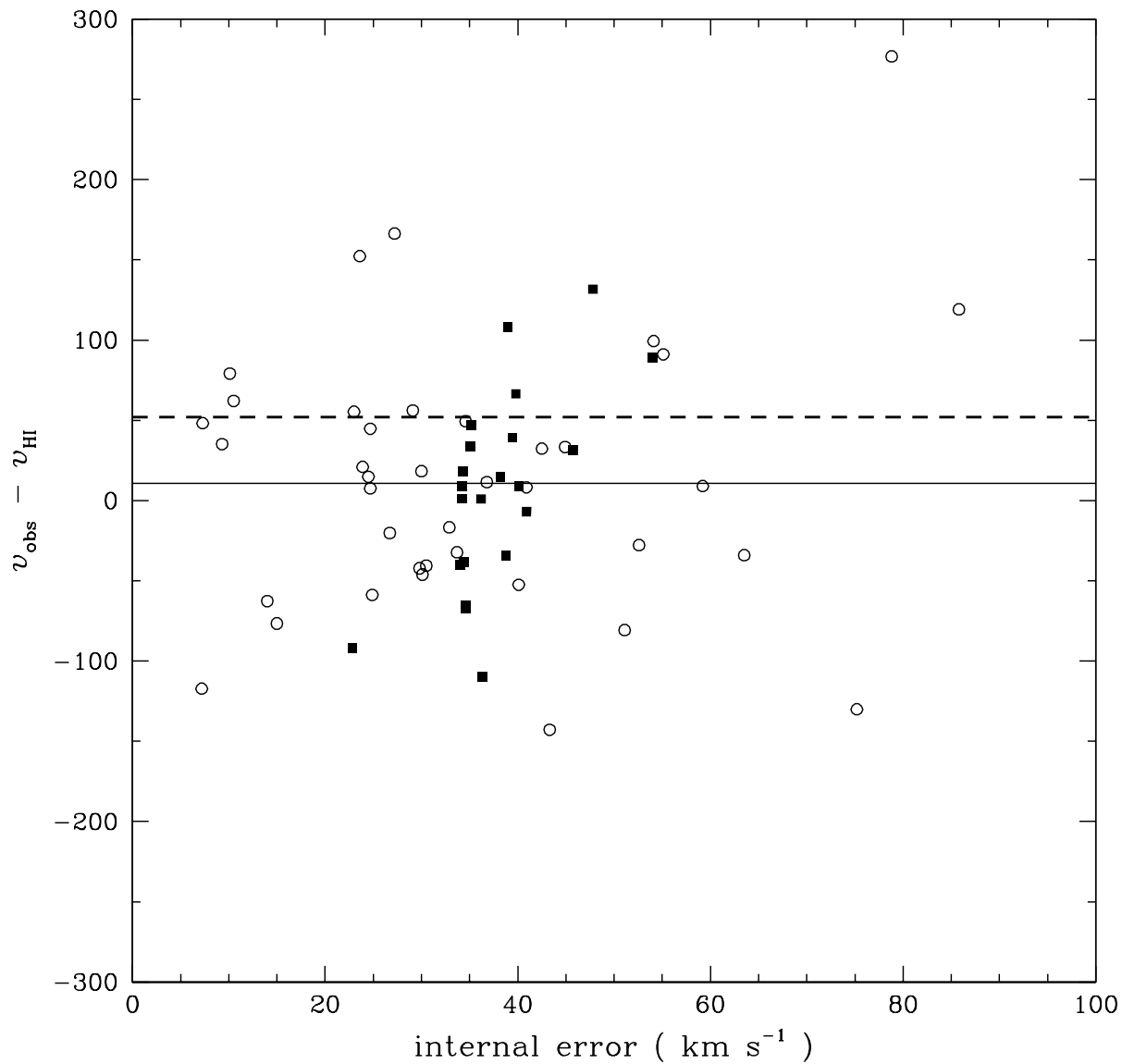}
\caption{Residual radial velocities of 61 galaxies relative to H I redshifts from NED as a function of internal velocity error. The residuals are consistent with a Gaussian distribution with $\sigma$=76 km s$^{-1}$. Filled boxes are galaxies in A1060, while open circles are galaxies from \citet{zabmul98}. The solid line shows the mean residual. The dashed line shows the mean residual of 336 stars observed with the same instrument, consistent with the distribution of galaxy residuals.}
\end{figure}

\subsection{Imaging}

We imaged each cluster in a mosaic of 25 tiles, which yielded a complete R-band photometric catalog of galaxies. The R-band photometry is both higher-resolution and less sensitive to recent star formation history than the ${b_{J}}$-band photometry obtained from DSS images. The R-band catalog defines the sample completeness of the spectroscopic survey. We discuss possible biases that may result from the fact that we used different magnitude bands for the target selection and the final photometric catalog in \S 3.1. We obtained the bulk of the R-band imaging data with the $40^{\prime\prime}$ Swope telescope and the TEK4 CCD at the Las Campanas Observatories. Some missing exposures were provided by Dennis Zaritsky with the same instrument and by Jose Arenas, using the SITe\#3 CCD at the same telescope. Typical seeing for most exposures was 1-2 arcseconds, with a pixel scale of 0.696 arcsec/pixel, sufficiently small to allow a robust star/galaxy separation down to $m_{R}\leq19$. We took most images under photometric conditions, but even the non-photometric images have a scatter in their photometric calibration of only a few hundredths of a magnitude.

The survey region around each cluster consists of a mosaic of 5$\times$5 tiles, covering an area of 1.5$\times$1.5 square degrees. The sampling radii in Mpc are also given in Table 1. The exposure time on each tile is 2$\times$120 s or, in a few cases, a single exposure of 240 s.

\subsection{Image processing}

We subject each frame to a bias subtraction and flatfielding using sky flats. Where both dome flats and sky flats are available, we use dome flats for the flatfielding and sky flats for a subsequent illumination correction. In a few cases where the standard flat fields are problematic, we create a flat field by combining a large number of object frames and rejecting high pixels to remove all objects. If more than one exposure is available for a given tile, we combine them by addition. 

We repair bad columns in the combined images using the IRAF fixpix task. We treat other artifacts, such as extended gradients emanating from a number of bad columns or pedestals in some regions of the chip, or scattered starlight from some of the brightest galactic foreground stars, using custom made algorithms that remove structures defined either by particular symmetries (such as circularly symmetric scattered light around stars or defects extending over entire columns) or by characteristically large scales of their brightness fluctuations. 

From the resulting image, we then construct our catalog of detections and perform photometry. In order to identify and remove cosmic rays from the catalog, we also produce a second combined image by averaging the two exposures (where available) with the minmax rejection algorithm set to reject the highest value among the two frames at each pixel. This procedure produces a combined frame that is virtually free of cosmic rays; we use this frame only to identify spurious detections that are likely to be cosmic rays, but do not extract photometry from it.

We use the Source Extractor software \citep{bertin} to construct a catalog of detections from each combined object frame and to perform photometry. An object is detected if at least 5 pixels lie above the threshold of 23.61 mag arcsec$^{-2}$, which corresponds to a deviation of an individual pixel of 1.5 - 2 $\sigma$ over the background on most tiles. Our surface brightness limit is comparable to that of \citet{sloanlf}, and the use of Source-Extractor total magnitudes avoids the biases that they find to be associated with the use of isophotal magnitudes with shallow surface brightness limits.

We use a combination of three different methods to separate stars and galaxies. The first is the `stellarity' flag provided in the Source-Extractor output. This classifier is based on a neural network algorithm. Our second method is based on the fact that galaxies and stars occupy two different, distinct regions in a plot of apparent magnitude versus surface brightness. For each frame, we visually determine a separatrix that optimizes the separation of stars and galaxies in this parameter space. The third method is visual inspection. We inspect every detection that is not unambiguously identified as a galaxy by the two automatic mechanisms, and classify it as stellar, galactic or uncertain. In cases of conflicts between the two automated methods, our visual classification tends to confirm the Source Extractor classification. Therefore, we count as galaxies any objects that we classify visually as galaxies, as well as those that we visually classify as uncertain, but that are classified as galaxies by Source Extractor.

 Details of this procedure and of the construction of our detection catalog in general are provided in Appendix A.

We calibrate the coordinate transformation between image and equatorial coordinates (2000.0) for our detections using the Guide Star Catalog-I \citep{gscI} and, in few cases, the Guide Star Catalog-II \citep{gscII}. For photometric calibration, we use standard star fields from \citet{graham} and \citet{landoldt}. The astrometric errors are on the order of the internal consistency of the GSC; we have not encountered any problems matching up detections with spectroscopic targets or with other detections of the same object in overlapping fields.

\subsection{Photometry}

For the total magnitude $m_{R}$, we adopt the photometric value provided by Source Extractor as $m_{best}$, except in the few instances described in Appendix A. This is an automatic estimate of the total magnitude of the object, determined either from an adaptive aperture magnitude or from a correction to the isophotal magnitude (see \citet{bertin}). For a few objects (not selected systematically), we have verified the Source Extractor magnitudes using a separate program that performs photometry on individual objects and find the agreement to be very good ({\it i.e.}, typically accurate to $<$0.05 mag) for reasonably bright, well-isolated bright objects.

After performing an initial photometric calibration of the magnitude zero point using the appropriate standard star fields for each night, we allow for a small photometric correction for each plate that minimizes the photometric discrepancies in the overlap regions between adjacent tiles (approximately $4^{\prime}$ wide). 

To determine that correction, we describe the quality of photometric agreement among the 25 tiles by
\begin{equation}
\Phi = \sum_{i}\sum_{j}(\delta_{ij}+\Delta_{i}-\Delta_{j})^{2}N_{ij}
\end{equation}
Here, $\delta_{ij}$ is the systematic magnitude difference between stars and galaxies on tiles $i$ and $j$. We must solve for $\Delta_{i}$, the magnitude correction of tile $i$ ({\it i.e.}, the systematic offset that is to be subtracted from all magnitude measurements in tile $i$). $\Phi$ is thus basically a least-squares estimator, weighted by the number $N_{ij}$ of matches found in the overlap regions.
By requiring that $\partial\Phi/\partial\Delta_{i}=0$, we find
\begin{equation}
\Delta_{i}=\frac{\sum_{j}N_{ij}(\Delta_{j}-\delta_{ij})}{\sum_{j}N_{ij}}
\end{equation} 
This equation is suitable for an iterative solution for the photometric correction $\Delta_{i}$. An additional constraint is that there should be no net magnitude offset over the entire 25 tiles (the results do not differ noticeably if that requirement is restricted to exposures taken on photometric nights).

Even on tiles imaged on different nights, the photometric corrections $\Delta_{i}$ are of the order of a few hundredths of a magnitude at most (the {\it rms} correction is $\sim$0.035), indicating that the photometric calibration is stable within our typical magnitude errors and that atmospheric extinction effects are minor.

We estimate the random errors remaining in our photometry by comparing the total magnitudes of galaxies in the overlap regions, and by determining the magnitude intervals within which about 2/3 of the comparisons agree. These intervals are dependent on apparent magnitude, and we approximate them by the following empirical results: $\Delta m=0.02$ at $m=13$, $\Delta m=0.04$ at $m=17$, $\Delta m=0.08$ at $m=18.5$, $\Delta m=0.11$ at $m=20$. When quoting magnitude errors for individual galaxies, we extrapolate the errors linearly between these data points. Note that these errors reflect only internal consistency.

In addition, we have compared our photometry to literature values in the NED database. For these comparisons, we have used total (preferred) or isophotoal magnitudes to surface brightness limits of 26 or 25 mag arcsec$^{-2}$ in the R-band. The magnitude discrepancies for individual galaxies typically scatter between +0 and +0.2 mag (our measurements typically yield the fainter values), although there are outliers at magnitude differences of about 0.4 mag. The distribution of magnitude differences, especially at fainter magnitudes, is non-random, presumably because of the nonhomogeneity of the literature sources.

 A systematic offset in the photometry provided by Source-Extractor has been reported before (Daniel McIntosh, 2001, private communication). \citet{bertin} also quote a possible offset as large as 0.06 mag at $m_{R}=17$. We do not make an attempt to correct for this, as it does not affect our luminosity functions, but this problem should be kept in mind when interpreting the magnitudes that we quote for bright galaxies.

We examine galaxies with large ($>$0.2 mag) magnitude discrepancies with previously published values by remeasuring their photometry as described in Appendix A. There is no indication of systematic errors affecting individual galaxies in our sample, revealing no need to revise our photometry for these objects. The two most difficult objects are NGC 3309 and NGC 3311, which are located in the center of A1060 and have extended, overlapping envelopes. We model these galaxies individually using the ELLIPSE and BMODEL tasks in IRAF. The magnitudes that we quote for these two galaxies are based on this flux measurement, not on Source-Extractor.

To correct our magnitudes and surface brightnesses for absorption from foreground Galactic dust, we use the all-sky dust maps and the conversion factor for the CTIO R-band by \citet{schlegel}.

\subsection{The Galaxy Catalog}

Table 2 presents a sample from our catalog, listing spectroscopically sampled galaxies with positions, redshifts, and $m_{R}$ (uncorrected for galactic extinction). The coordinates listed are the target coordinates for the spectrographic fiber; in cases where the spectroscopic and photometric coordinates deviate by more than $3^{\prime\prime}$ ({\it e.g.}, because of confusion with a nearby star), we list the latter coordinates as a comment. The comments also note whether the object has been deblended from another detection (``deblend'') and/or manually added to the catalog (``add''). The comment ``mag!'' indicates that the apparent magnitude has been changed from the default Source Extractor value.

\begin{deluxetable}{lllllllll}
\rotate
\tabletypesize{\scriptsize}
\tablecaption{Galaxy Catalog (Example)}
\tablewidth{0pt}
\tablehead{
\colhead{ID}&\colhead{RA (J2000)}&\colhead{Dec (J2000)}&\colhead{$m_{R}$}&\colhead{$\Delta m_{R}$}&\colhead{cz}&\colhead{$\Delta cz$}&\colhead{NED ID}&\colhead{Comments}}
\startdata
  1060C\_494[104] & 10 36 42.70 & -27 31 42.00 & 10.07 & 0.015 &  3857 &  80 & NGC 3311                       & 10:36:43.21  -27:31:32.09   mag! add    \\
  1060A\_494[103] & 10 36 35.69 & -27 31  5.30 & 10.74 & 0.015 &  4071 &  80 & NGC 3309                       &  mag! add deblend                       \\
   1060A\_494[68] & 10 37  2.53 & -27 33 53.60 & 11.34 & 0.015 &  2761 &  80 & NGC 3312                       &                                         \\
   1060A\_494[91] & 10 37 47.30 & -27  4 52.00 & 11.46 & 0.015 &  2973 &  80 & IC 2597                        &                                         \\
   1060A\_494[33] & 10 33 35.60 & -27 27 17.20 & 11.54 & 0.015 &  3295 &  80 & NGC 3285                       &                                         \\
   1060A\_494[96] & 10 36 22.31 & -27 26 17.50 & 11.77 & 0.015 &  3537 &  80 & NGC 3308                       &                                         \\
   1060B\_494[22] & 10 36 57.88 & -28 10 38.80 & 12.24 & 0.015 &  2503 &  80 & ESO 437- G 015                 & 10:36:58.02  -28:10:35.58               \\
   1060B\_494[69] & 10 37 37.26 & -27 35 38.50 & 12.36 & 0.015 &  3922 &  80 & NGC 3316                       &                                         \\
    1060B\_494[3] & 10 36 12.04 & -27  9 43.20 & 12.43 & 0.015 &  4002 &  80 & NGC 3305                       & 10:36:11.74  -27: 9:43.92               \\
   1060A\_494[47] & 10 33 30.14 & -26 53 50.10 & 12.62 & 0.015 &  3535 &  80 & ESO 501- G 013                 &                                         \\
   1060A\_494[70] & 10 37 12.76 & -27 41  1.10 & 12.62 & 0.015 &  2795 &  80 & NGC 3314                       &                                         \\
   1060A\_494[78] & 10 37 19.17 & -27 11 30.50 & 12.75 & 0.015 &  3753 &  80 & NGC 3315                       &                                         \\
   1060B\_494[25] & 10 36 50.43 & -27 55  8.80 & 12.78 & 0.015 &  4854 &  80 & ESO 437- G 011                 &                                         \\
   1060A\_494[39] & 10 34 36.75 & -27 39  9.30 & 12.90 & 0.015 &  3150 &  80 & NGC 3285B                      & 10:34:36.97  -27:39:10.25               \\
   1060A\_494[17] & 10 36 24.72 & -26 59 57.60 & 12.97 & 0.015 &  4115 &  80 & ESO 501- G 035                 &                                         \\
   1060A\_494[20] & 10 36 53.99 & -27 54 58.90 & 13.08 & 0.015 &  3625 &  80 & ESO 437- G 013                 & 10:36:53.98  -27:55: 2.16               \\
   1060A\_494[57] & 10 38 33.32 & -27 44 12.40 & 13.12 & 0.016 &  4412 &  80 & ESO 501- G 065                 &                                         \\
   1060A\_494[64] & 10 39 18.26 & -26 50 23.50 & 13.16 & 0.016 &  3113 &  80 & ESO 501- G 068                 &                                         \\
    1060A\_494[6] & 10 35 20.48 & -27 21 42.90 & 13.19 & 0.016 &  4539 &  80 & ESO 501- G 021                 &                                         \\
   1060A\_494[75] & 10 37  4.89 & -27 23 59.30 & 13.28 & 0.016 &  2690 &  80 & PGC 031515                     &  deblend                                \\
   1060A\_494[24] & 10 36 32.45 & -28  3 48.90 & 13.29 & 0.016 &  4362 &  80 & ESO 437- G 008                 &                                         \\
   1060A\_494[92] & 10 37 49.40 & -27  7 15.20 & 13.29 & 0.016 &  2434 &  80 & ESO 501- G 059                 &                                         \\
   1060A\_494[30] & 10 34 47.70 & -27 12 51.50 & 13.31 & 0.017 &  4369 &  80 & ESO 501- G 020                 &                                         \\
   1060A\_494[77] & 10 37 17.01 & -27 28  7.60 & 13.36 & 0.017 &  4821 &  80 & ESO 501- G 047                 &                                         \\
  1060A\_494[100] & 10 36 31.80 & -27 13 14.90 & 13.46 & 0.017 & 10689 &  80 & [RMH82] 30                     &                                         \\
   1060A\_494[23] & 10 36 34.61 & -28 12 52.80 & 13.48 & 0.017 &  3652 &  80 & ESO 437- G 009                 &                                         \\
   1060A\_494[98] & 10 36 27.64 & -27 19  8.50 & 13.48 & 0.017 &  3376 &  80 & PGC 031447                     &                                         \\
   1060A\_494[48] & 10 34 59.59 & -28  4 41.80 & 13.55 & 0.018 &  2466 &  80 & ESO 437- G 002                 &                                         \\
   1060A\_494[84] & 10 37  5.00 & -27 59  9.60 & 13.57 & 0.018 &  3984 &  80 & PGC 031517                     &                                         \\
  1060A\_494[105] & 10 36 41.15 & -27 33 39.10 & 13.69 & 0.018 &  4735 &  80 & SGC 1034.3-2718                &                                         \\
   1060A\_494[60] & 10 39 24.92 & -27 54 46.20 & 13.73 & 0.019 &  3196 &  80 & ESO 437- G 032                 &                                         \\
  1060A\_494[101] & 10 36 44.89 & -27 28  9.80 & 13.75 & 0.019 &  2735 &  80 & PGC 031483                     &                                         \\
   1060A\_494[71] & 10 37 19.95 & -27 33 33.70 & 13.81 & 0.019 &  4020 &  80 & ESO 501- G 049                 &                                         \\
   1060A\_494[11] & 10 36 17.12 & -27 31 46.50 & 13.84 & 0.019 &  3773 &  80 & NGC 3307                       &                                         \\
   1060A\_494[31] & 10 33 59.72 & -27 27  5.70 & 13.86 & 0.019 &  3545 &  80 & ABELL 1060:[R89] 129           &                                         \\
\enddata

[The complete version of this table is in available at URL http://hotspur.as.arizona.edu/{$\sim$}dchristl/tab2.txt]
\end{deluxetable}

From this spectroscopic sample, we select various subsamples. For each of the six cluster fields, the cluster members lie in the redshift ranges in Table 1. The members are further split into emission line (EL) and non-emission line (NEL) samples. EL galaxies are those with $\lambda3727$ [OII] doublet equivalent widths $>5$\AA; NEL galaxies are the rest. Thus, EL galaxies represent star forming or active galaxies, and NEL galaxies are relatively quiescent.

We also create a composite sample of all cluster galaxies, as well as a composite sample of field galaxies. We define the field sample as all galaxies not explicitly included in any of the cluster samples. As the redshift space in the direction of several of these clusters reveals large-scale structures at different redshifts, the field sample actually represents a range of environments and is not restricted to truly isolated galaxies. A NED database search shows no other major clusters within 1 degree of our lines of sight, but there are several clusters at slightly larger projected distances, indicating that the field sample includes higher density regions as well.

We further complement our catalog with redshift data taken from NED (this adds 86 redshifts to the sample of 1563 cluster galaxies and 46 redshifts to the sample of 703 field galaxies). The inhomogeneity of the literature sources raises concerns that they may be biased towards cluster members, but the overall contributions to our sample are small enough not to constitute a problem. These additional redshifts are used only for the calculation and analysis of the overall GLFs, not for the EL and NEL subsamples, because [OII] equivalent widths are typically not available for the objects supplemented from NED.

\section{Calculating the Luminosity Function}

\subsection{The Sampling Fraction}

In any statistical investigation of galaxy properties, each galaxy has to be weighted by the inverse of the sampling fraction $f_{s}$, which is the fraction of photometrically detected galaxies obeying certain selection criteria ({\it e.g.}, cluster membership, luminosity) that have been spectroscopically sampled.

Due to the design of this survey (target selection by apparent magnitude, multiple spectroscopic exposures of each field), the sampling fraction is not dependent on the position of an object on the sky or the proximity of other targets. The primary dependence of $f_{s}$ is on apparent magnitude, and possibly on surface brightness. It is possible that systematic discrepancies exist in the star/galaxy classification between the initial target selection (made from lower resolution Digitized Sky Survey (DSS) plates) and the final photometric catalog (made from CCD images as described earlier), but we expect these to be correlated with the position of a galaxy detection in the $(m_{R};\mu_{R})$ plane.

We therefore choose to calculate the sampling fraction by counting photometrically detected and spectroscopically sampled galaxies as a function of $(m_{R},\mu_{R})$ in overlapping bins of a fixed size on a fine grid in the $(m_{R};\mu_{R})$ plane. We thus obtain a (smoothed) estimate of the sampling fraction at every point in the plane where galaxies have been detected.

We inspect visually all objects not identified as stars by both algorithms described in Appendix A. We assign full statistical weight to visually confirmed galaxies, as well as to visually uncertain objects classified as galaxies by Source-Extractor. To be conservative, we also visually inspect objects identified as stellar by both algorithms in up to a third of the mosaic tiles per cluster. We reclassify a small fraction ($<<1\%$) of these objects as galaxies and assign this fraction as a fractional statistical weight to the uninspected stellar objects with similar $(m_{R};\mu_{R})$ on the remaining tiles. All other ({\it i.e.} non-stellar) detections that we do not confirm visually are presumed spurious and discarded.

The spectroscopic sample covers a region of $(m;\mu)$ space in which galaxies are unambiguously identified, so that most of the detections used in the calculation of $f_{s}$ are unambiguous galaxies. To assess the impact of the star/galaxy classification on our results, we consider two extremes: objects classified as galaxies by all three methods and objects classified as non-stellar by at least one method. Between these two extremes, the faint end slope $\alpha$ of the GLF changes only by $\sim$0.05, and our default option yields results in the middle of this interval.

Fig. 2 shows the $(m_{R};\mu_{R})$ plane and the average sampling fraction (including the redshifts added from NED) for all six fields. The distribution of all galaxy detections in our catalog is bounded by the light solid line \footnote{Note that the sampling in individual fields may vary from this averaged distribution; our analysis therefore treats the sampling fraction as a function of $m_{R}$, $\mu_{R}$, and cluster field}. The figure also shows the region of $(m_{R};\mu_{R})$ represented by the spectroscopic sample (bold envelope). The standard magnitude limit for our composite GLFs is $m_{R}=18$. As we have no information on the redshift distribution of galaxies outside the spectroscopically sampled region, those detections do not contribute to our GLFs. It is obvious from Fig. 2 that such regions are very small; the only population not sampled is one of faint high-surface brightness galaxies that appeared stellar on the lower-resolution DSS plate material and thus were not targetted. Most are fainter than our magnitude threshold of $m_{R}=18$.

\begin{figure}
\plotone{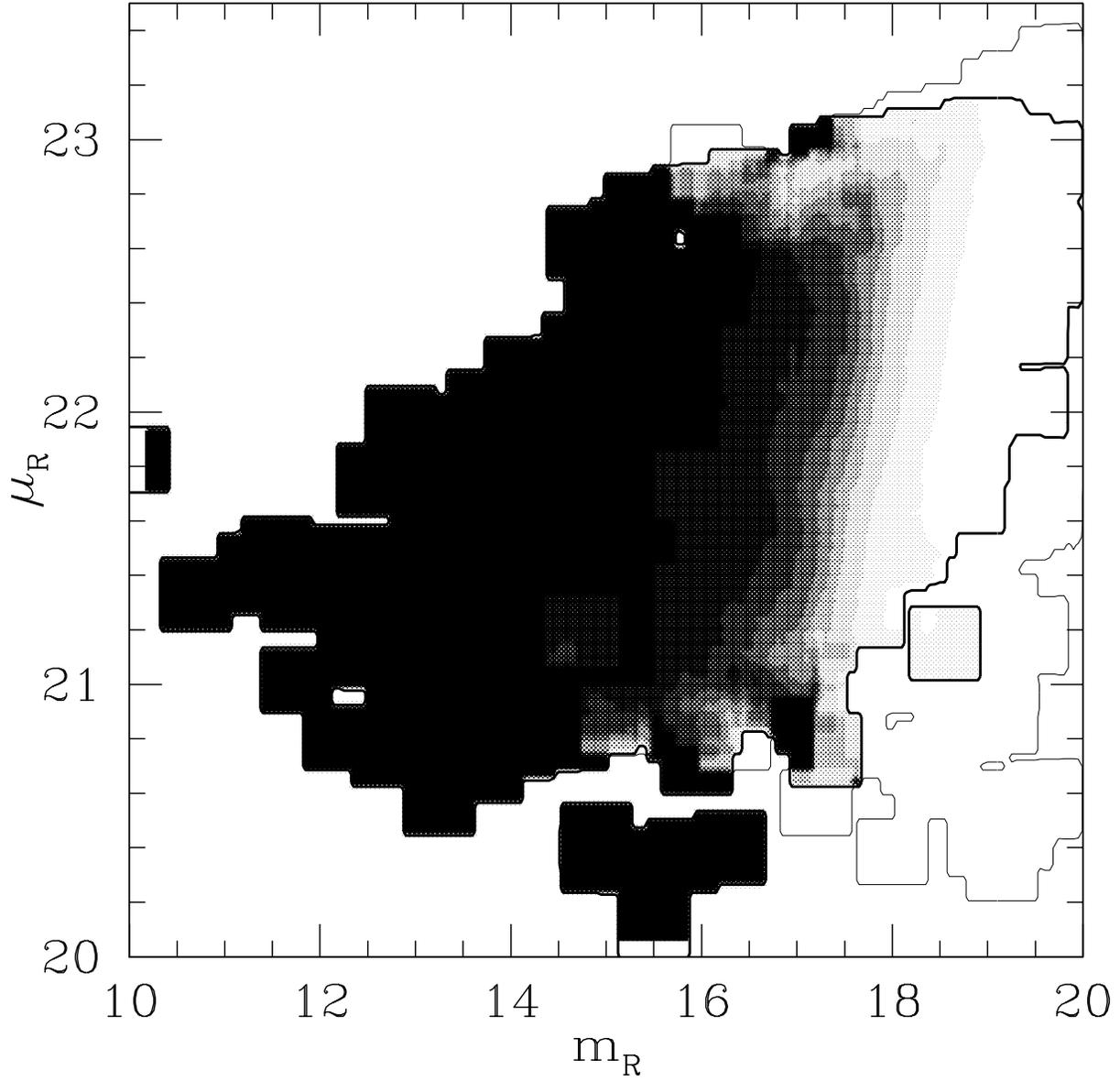}
\caption{Averaged galaxy sampling fraction as a function of $(m_{R};\mu_{R})$. Greyscales indicate the spectroscopic sampling fraction, ranging from 0 to 1. The bold envelope denotes the spectroscopically sampled region ({i.e.}, all $(m_{R};\mu_{R})$ for which $f_{s}>0$); the light envelope denotes the regions containing photometric detections.}
\end{figure}

We also note a dependence of the sampling fraction on surface brightess at faint magnitudes. The sampling fraction drops strongly at low-surface brightness end of the distribution, but also exhibits a decline towards the high-surface brightness end. We have randomly looked up several objects in this undersampled region that are clearly galaxies on our R-band exposures; these objects appear faint and star-like on the DSS plates, with no discernible diffuse component. We conclude that the reason for the decline in the sampling fraction with increasing surface brightness lies in the initial star/galaxy separation during the target selection.

A few galaxies stand out in this distribution at unusually high surface brightness values. For these, we find no evidence of errors in the photometry or visual classification. One of these objects with spectroscopic data, 1631A\_494[13], is very compact, but has a nonstellar profile. This galaxy is superimposed on the envelope of NGC 4756, with a redshift difference of more than 300 km/s. The large difference in surface brightness between the object and the envelope of NGC 4756 makes a confusion of the spectra unlikely. Another object, SERA\_1294[33] (APMUKS(BJ) B042429.42-611817.9), has a visible disk, albeit of very low surface brightness, and a bright, compact bulge. Neither of these objects has unusual activity in its spectrum.

The analysis of the GLF should not be based on detections from poorly sampled regions of $(m_{R};\mu_{R})$ space, where possible systematic errors associated with the sampling fraction might become significant. This is particularly important for the composite GLFs, where systematic deviations in the GLF at faint apparent magnitudes could manifest themselves as systematic deviations of the GLF over a wide range of absolute magnitudes. We therefore introduce an apparent magnitude limit for our analysis. We also quantify below some of the major effects that could bias the sampling fraction at faint magnitudes.

For the calculation of the composite GLFs, we truncate our catalog at an apparent magnitude of $m_{R}=18$ for five of our clusters. At this magnitude, the fractional completeness has dropped to $\sim$20-50$\%$ for the individual clusters. (Over the entire magnitude range down to that limit, the sample is much more complete on average, reaching nearly 100\% completeness for $m_{R}<16$). An exception is the field of A3266, where the sampling is shallower. Here we truncate at $m_{R}=17$ to achieve a similar level of completeness (although extending this limit to $m_{R}=18$ makes no significant difference in the GLF). We have also experimented with modeling the sampling fraction as a function of U-band magnitudes, which have been provided for three of our clusters by Daniel McIntosh. After correcting the galaxy counts by $f_{S}(m_{U};\mu_{R})$, we find that residual incompleteness remains for galaxies with $m_{R}>18$, indicating that color terms are important for fainter magnitudes. This provides an additional justification for cutting our sample at $m_{R}=18$.

For the intercomparisons of our GLFs, we adopt similar magnitude limits for reasons of consistency. In particular, when comparing clusters to composite GLFs, we adopt a magnitude limit of $m_{R}=18$ for all six clusters. For A3266, this is fainter than the limit to which it contributes to the composite GLF, but the differences are insignificant both for the calculation of the GLF and for comparisons between the composite GLFs and A3266. We also choose a standard {\it absolute} magnitude limit for some of our analyses in order to provide a common baseline for all comparisons, rather than the maximum baseline permitted by the apparent magnitude limit. This absolute magnitude limit is $M_{R}=-18.35$ (-18.5 in binned distributions), which corresponds to the standard apparent magnitude limit of $m_{R}=18$ at the distance of the furthest cluster, A3266.

A potential source of bias is the fact that target selection for this survey is based on approximate $b_{J}$ magnitudes, while our photometric catalog (and thus our calculation of the sampling fraction) uses R-band magnitudes. Cluster galaxies may have colors systematically different than field galaxies. A cluster galaxy of a given $m_{R}$ would thus have a different probability of being selected as a spectroscopic target than a field galaxy with the same $m_{R}$. Our sampling fraction, which is based on total counts of galaxies, irrespective of their cluster membership, would then be biased. By assuming a mean color difference between field and cluster galaxies, it is possible to reconstruct the field- and cluster-specific sampling fractions. We use these to estimate the magnitude of this effect and find that it is negligible. Appendix B gives the details of our calculation. 

Another potential bias is that our spectroscopic success rate may be correlated with the spectral properties of the target galaxies and thus indirectly with cluster membership. It is not obvious if this bias would favor the sampling of galaxies in clusters (because of the prevalence of extreme early types with strong absorption features) or in the field (because of the prevalence of extreme late types with strong emission lines). However, the number of spectroscopic targets for which we could not obtain a redshift provides a constraint on the magnitude of any such uncertainty (presuming that the targetting itself is representative). The upper and lower limits on the sampling fraction due to these failed targets show that the impact on our results is minor.

Our procedure for determining $f_{S}$ by counting discrete detections in $(m_{R};\mu_{R})$ bins of finite size is also subject to several biases. We choose bin sizes to minimize these effects, although a compromise between good statistics and an accurate representation of the sampling fraction is necessary when a smooth, but non-analytic function such as $f_{S}(m_{R};\mu_{R})$ is probed only at discrete points. We explain the details below.

A small bin size is likely to exclude photometric detections in regions with sparse spectroscopic sampling; photometric detections that do not fall into a bin with at least one spectroscopically confirmed galaxy are effectively discarded, while the sampling fraction for galaxies that fall in that bin may be overestimated. Of course, certain populations of galaxies in the $(m_{R};\mu_{R})$ plane simply may not have been sampled. These detections should not be included in the calculation of the luminosity function anyway, as their redshift distribution is unknown. We must therefore choose a $(m_{R};\mu_{R})$ bin size such that a spectroscopically smapled galaxy is representative of photometric detections in the same region.

A different problem lies in the fact that large bin sizes smooth over variations in the detection density and in the sampling fraction itself across the bin. Higher-order variations in the density of photometric detections (or spectroscopic galaxies) mean that the integrated number of galaxies across the bin is not representative of the detection density at the bin center.  We present a simple procedure to correct this effect to first order in $m_{R}$ in Appendix C, but use it only to quantify the magnitude of the effect, which is negligible.

There may also be small-scale variations of the sampling fraction itself in the $(m_{R};\mu_{R})$ plane; in particular, the extreme high and low surface brightness regions of the galaxy distribution are undersampled. A large bin size will smooth over these variations, overestimating the sampling fraction in the undersampled regions and underestimating it in the well-sampled regions. The effect on the luminosity function is dependent on how the population of the selected sample --- field or cluster galaxies --- is distributed in the $(m_{R};\mu_{R})$ plane. 

Our choice of the bin size for the calculation of the sampling fraction in $(m_{R};\mu_{R})$ is motivated by the robustness of the GLF that it produces. The faint end slope of the field GLF is particularly sensitive to the second of the aforementioned effects, the variation of the detection density across the bin, and thus places constraints on the choice of the bin size in $m$. A bin size of $\Delta m=0.75$ is small enough to avoid these higher-order effects, while not running into the problem of excluding too many photometric detections in sparsely sampled regions. For this bin size, $\Delta\mu\:=\:0.25$ mag arcsec$^{-2}$ yields robust results for the GLF.

\subsection{The Individual Cluster GLFs}

We calculate the cluster GLFs using bins in $M_{R}$ of width $\Delta M_{R}=0.5$ (adopted as a compromise between sufficiently high signal-to-noise in most bins and sufficient resolution of the shape of the GLF). We add up the number of galaxies in each bin, weighted by the inverse of their sampling fraction. This is a justifiable procedure for clusters, as all galaxies are at approximately the same distance and thus do not require a volume correction.

The finite extent of the lowest redshift cluster in the sample, A1060, leads to a small additional uncertainty in the distance modulus and thus the absolute magnitude of each galaxy. For A1060 and $H_{0}=100$ km s$^{-1}$ Mpc$^{-1}$, this uncertainty is on the order of 0.07 mag, much smaller than our bin size.

\subsection{The Field and Composite Cluster GLFs}

For calculating the field GLF (and the composite cluster GLF), it is necessary to adopt a different approach, as the observation of galaxies over a wide range of distances requires a volume correction (more luminous galaxies can be seen over larger distances and are thus overrepresented by numbers in a magnitude-limited sample). Weighting by the inverse volume over which a galaxy would be visible would be unlikely to yield good results, as it assumes galaxies are distributed homogeneously in comoving space.

We use a stepwise maximum likelihood (SWML) estimator as described by \citet{eep}. The derivation below is modelled after that given in \citet{linetal}, with modifications to account for the fact that, instead of fixed apparent magnitude limits, we have a variable sampling fraction as a function of $(m;\mu)$. Each galaxy $i$ in the sample is characterized by an absolute magnitude $M_{i}$ and an ``absolute surface brightness'' $\mu_{i}$ ({\it i.e.}, before cosmological effects or Galactic extinction are applied). Also, each galaxy is associated with a particular redshift, Galactic extinction, and one of our six cluster fields. We parametrize these last three variables, which determine the sampling probability of a galaxy with a given absolute magnitude and surface brightness, by the vector $F_{i}$. The probability that a sampled galaxy with the properties $F_{i}$ has the absolute magnitude $M_{i}$ and surface brightness $\mu_{i}$ is

\begin{equation}
p_{i} = p(M_{i};\mu_{i}\mid F_{i}) = \Phi(M_{i},\mu_{i}) * f_{s}(M_{i};\mu_{i};F_{i}) / \int \Phi(M;\mu) f_{s}(M;\mu;F_{i}) dM d\mu
\end{equation}

Note that that $f_{s}(M_{i};\mu_{i};F_{i})$ is primarily a function of apparent magnitude and surface brightness and may also vary from one cluster field to another. Given $F_{i}$, the conversion between absolute and apparent variables is unambiguous. The integral is over all of $(M;\mu)$ space.

The logarithmic likelihood function for a sample consisting of $N$ galaxies is then
\begin{equation}
ln {\cal L} = \sum_{i=1}^{N}[ln(\Phi(M_{i},\mu_{i})*f_{s}(M_{i};\mu_{i};F_{i})) - 
ln \int \Phi(M;\mu) f_{s}(M;\mu;F_{i}) dM d\mu]
\end{equation}

We then discretize $\Phi$ in $(M;\mu)$ bins, writing $\Phi_{kl}$ for $\Phi(M_{k};\mu_{l})$:
\begin{eqnarray}
ln {\cal L} = \sum_{i}^{N} \sum_{k,l} ln(\Phi_{kl}*f_{s}(M_{k};\mu_{l};F_{i}))*W(M_{i};\mu_{i};M_{k};mu_{k})
- \\
\sum_{i}^{N} ln [\sum_{k,l} \Phi_{kl} f_{s}(M_{k};\mu_{l};F_{i}) * H(M_{k};\mu_{l};F_{i}) * \Delta M \Delta\mu]
\end{eqnarray}
where $W$ is $1$ if galaxy $i$ falls into bin $(k;l)$ and $0$ otherwise. $H$ contains fractional corrections to the bin widths to account for any overlap of the bin with the defined limiting magnitudes of the catalog (particularly the imposed apparent magnitude cutoff at $m_{R}=18$).
We take the derivative by $\Phi_{mn}$:
\begin{equation}
\frac{\partial ln \cal{L}}{\partial \Phi_{mn}} = \sum_{i} \Phi_{mn} W(M_{i};\mu_{i};M_{m};\mu_{n}) 
- \sum_{i} \frac{f_{s}(M_{m};\mu_{n};F_{i}) H(M_{m};\mu_{n};F_{i}) \Delta M \Delta\mu}
{\sum_{k,l} \Phi_{kl} f_{s}(M_{k};\mu_{l};F_{i}) H(M_{k};\mu_{l};F_{i}) \Delta M \Delta\mu}
\end{equation}
Now we set this expression to zero and solve for $\Phi_{mn}$ to obtain a prescription for an iterative solution:
\begin{equation}
\Phi_{mn} = \sum_{i} W(M_{i};\mu_{i};M_{m};\mu_{n}) / \sum_{i} \frac{f_{s}(M_{m};\mu_{n};F_{i}) H(M_{m};\mu_{n};F_{i}) 
\Delta M \Delta\mu} {\sum_{k,l} \Phi_{kl} f_{s}(M_{k};\mu_{l};F_{i}) H(M_{k};\mu_{l};F_{i}) \Delta M \Delta\mu}
\end{equation}

This maximum likelihood estimator converges fairly quickly. It has the advantage of being unbiased by large-scale structure inhomgeneities, as the sums implicitly trace the redshift distribution of the sample. Its application to clusters has the additional advantage of extracting information even from empty bins, which would be ignored by simply averaging individual GLFs.

When the sampling fraction for a given bin in $(M_{R};\mu_{R})$ is referenced by the algorithm, we calculate an average sampling fraction over the area of that bin. In some cases, the algorithm may attempt to reference the sampling fraction at coordinates in the $(m;\mu)$ plane where no galaxies have been sampled ({\it e.g.}, to determine the hypothetical visibility of a high-redshift giant galaxy if it were located in a low-redshift field). Therefore, we extrapolate sampling fractions in those unsampled bins prior to calculating the GLF with an iterative algorithm similar to the Liebmann method for solving the Poisson equation on a discrete grid. The effect of this approach on our GLFs, as compared to setting such undefined sampling fractions to zero, is smaller than our quoted uncertainties by more than an order of magnitude.

We do not apply a maximum likelihood estimator \citep{sty} to calculate parametric luminosity functions (such as Schechter (1976) functions), as there is no standard analytical expression that would allow us to model the galaxy distribution in luminosity and surface brightness simultaneously (but see \citet{crossanddriver} for a proposal for such a two-dimensional analogue to the Schechter function). Instead, we fit Schechter functions to the results of the stepwise maximum likelihood estimator by $\chi^{2}$ minimization (after they have been integrated over surface brightness) and estimate the errors in this procedure using the prescription by \citet{avni}.

The maximum likelihood method leaves the normalization of the composite GLFs undetermined, so that they are actually luminosity distributions. Nevertheless, we refer to them as GLFs throughout for reasons of simplicity.

To test the SWML algorithm, we apply it to individual clusters and compare the results to GLFs obtained from the direct binning method described in \S 3.2. Despite the slightly different treatment of the sampling fraction (for the direct binning method, $f_{s}$ is calculated in bins centered on the individual detections, rather than on a regular grid), the GLFs obtained from these two methods are indistinguishable. Furthermore, the field GLFs from the six individual fields are all consistent with the composite field GLF, as would be expected from a sample of similar environments, indicating that our algorithm reproduces the same field GLF even in fields with different redshift distributions and sampling fractions.

\section{Results and Discussion}

\subsection{Individual Cluster GLFs}

The individual luminosity functions for each of the six clusters are shown in Fig. 3 for all spectral types and in Figs. 4 and 5 for the emission line (EL) and non-emission line (NEL) subsamples, respectively. The error bars denote the Poisson errors of the spectroscopically sampled galaxies, modified by the sampling fraction. The thin solid lines correspond to the uncertainties in each GLF --- overall, EL, NEL --- due to the failed spectroscopic targets. For example, in the case of the overall GLF, the upper limit assumes that all failed spectroscopic targets are cluster members, and the lower limit that none of them are. We note that these uncertainties are small (typically within the Poisson errors), so that any correlation of the spectroscopic success rate with cluster membership or spectral properties cannot bias our results significantly. The dashed lines represent the most pessimistic scenarios where {\it all} unsampled galaxies are cluster members (upper limit) or not (lower limit).

\begin{figure}
\plotone{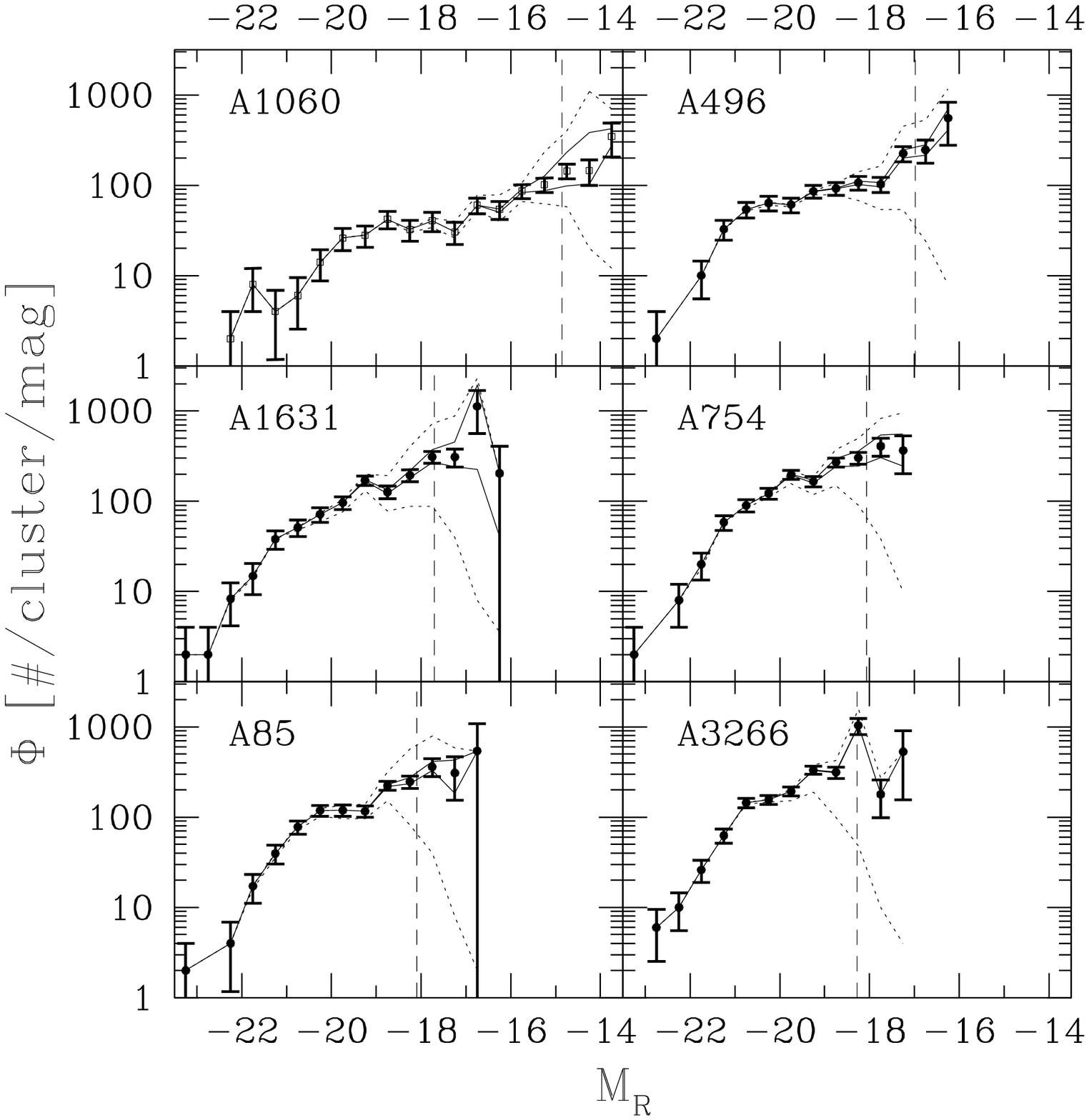}
\caption{Cluster GLFs for all spectral types. Clusters are arranged left to right and top to botton in order of increasing redshift. Error bars are $1\sigma$. Solid lines give upper and lower limits on the GLF, assuming that all spectroscopic targets for which we could not obtain redshifts are either members (upper limit) or non-members (lower limit). Dashed lines give number of spectroscopically confirmed sample members (lower limit) and total number of detections that are not confirmed non-members (upper limit). For orientation, vertical lines indicate the absolute magnitude corresponding to our standard analysis threshold of $m_{R}=18$.}
\end{figure}

\begin{figure}
\plotone{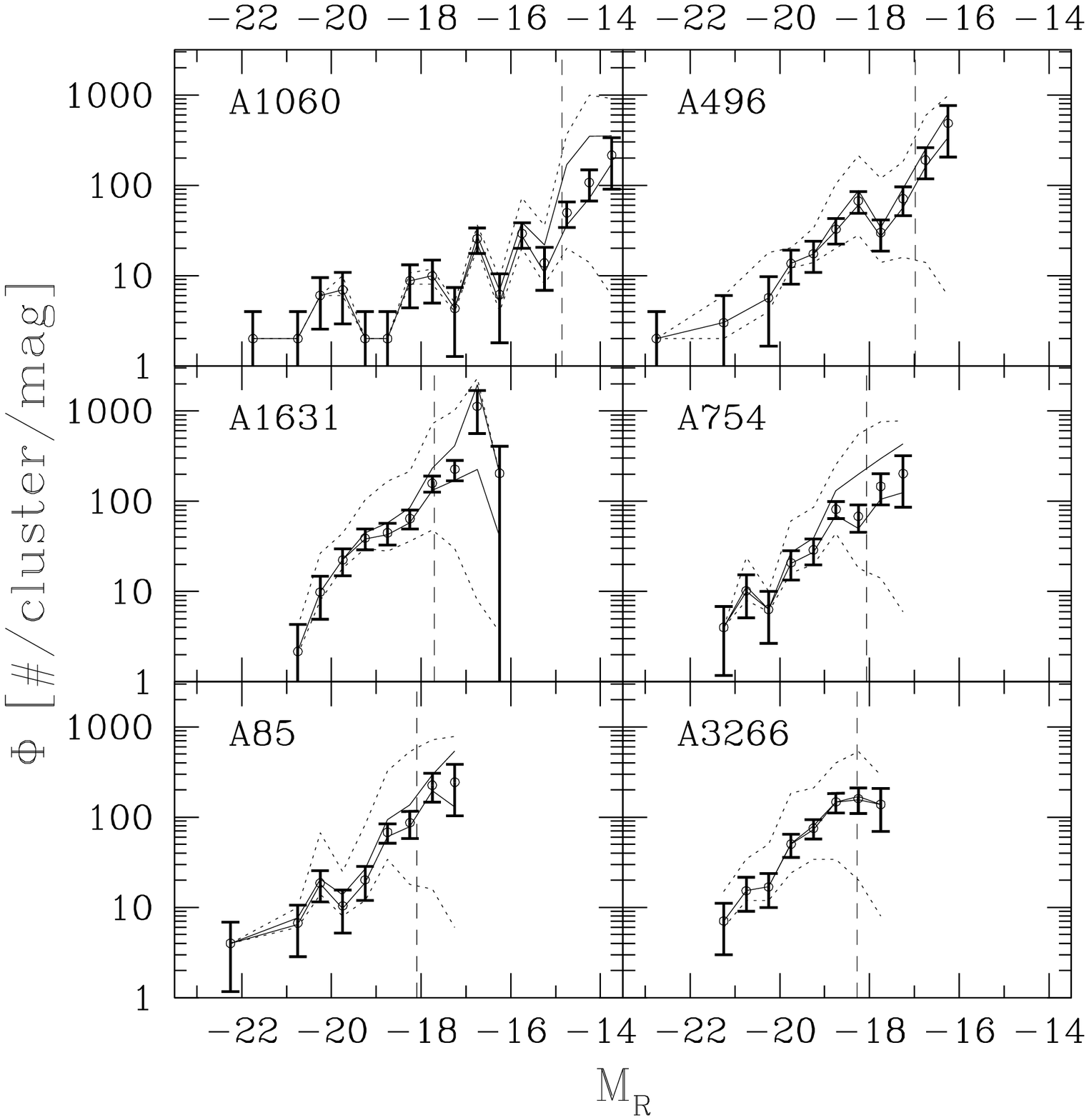}
\caption{Same as Fig. 3 for EL galaxies only.}
\end{figure}

\begin{figure}
\plotone{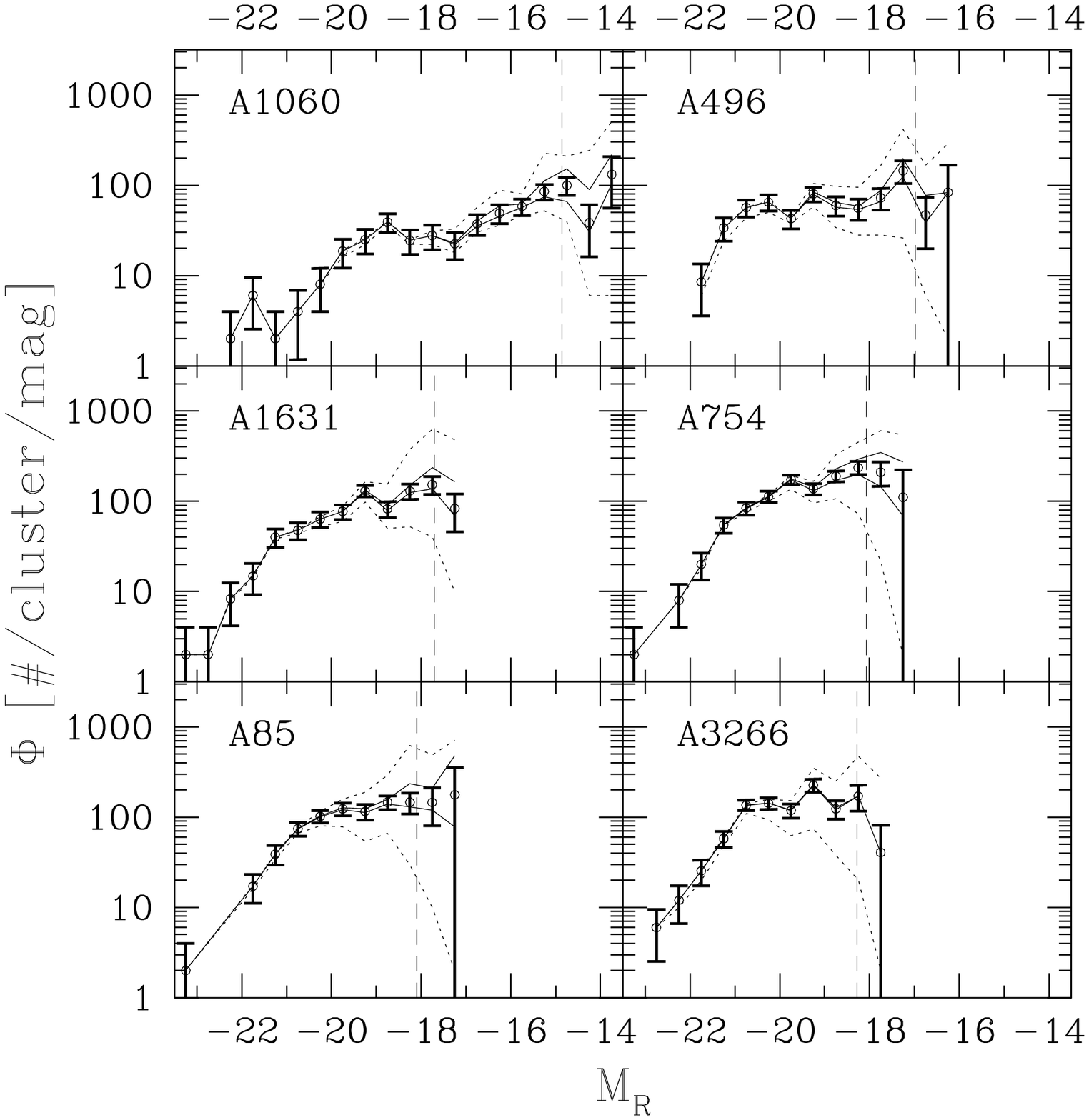}
\caption{Same as Fig. 3 for NEL galaxies only.}
\end{figure}

\subsubsection{Comparisons among Cluster GLFs}

We use two different comparison tests to examine whether the GLFs in our six clusters agree to within the statistical uncertainties. First, we compare the individual cluster GLFs to each other using a $\chi^{2}$ test. For each of the overall, EL and NEL samples, this test yields 15 independent probabilities. We then analyze the results of these 15 individual comparisons statistically to determine whether they are consistent with the hypothesis that differences between the six clusters are only random, and not systematic.

While this procedure will tell us whether our data is consistent with a universal cluster GLF or not, it does not yield a quantitative estimate of the ``universality'' of the GLF in clusters ({\it i.e.}, the fraction of all clusters for which the average GLF is representative). A constraint on this universality is useful in estimating how representative the composite GLFs that we present in \S 4.2 are for clusters in general. Therefore, we perform a second comparison test, in which we compare the individual clusters to composite GLFs. This test shows how many, and which, clusters in our sample are consistent with the composites, and this in turn allows us to constrain the universality of the composite GLFs. The details of both tests are below. 

We turn first to the cluster-to-cluster comparisons. In order to be consistent with our treatment of the composite GLFs, which are presented below, we adopt the same magnitude limits as discussed in \S 3.1. We impose an apparent magnitude cutoff of $m_{R}=18$ and compare the clusters only over absolute magnitude ranges brighter than this limit. Down to this apparent magnitude, we consider our sampling fractions reliable. For each of the EL, NEL and overall samples, we perform 15 cluster-to-cluster comparisons. If no systematic differences exist between clusters, we would expect about one in 20 comparisons to be inconsistent at the 2$\sigma$ level if the variance among clusters obeys Gaussian statistics.

Among the overall GLFs, only those of A1060 and A496 ($M_{R}\leq-17.5$) are inconsistent. There are no significant inconsistencies between the EL GLFs. In the NEL sample, we find inconsistencies between A1060 and A496  ($M_{R}\leq-17.5$), A496 and A754 NEL ($M_{R}\leq-18.5$), and A754 and A3266 NEL ($M_{R}\leq-18.5$).

To perform the same analysis over a common absolute magnitude range, we adopt a  absolute magnitude limit of $M_{R}=-18.35$ (effectively $-18.5$ due to the bin boundaries), approximately the limit that corresponds to $m_{R}=18$ for the most distant cluster. All individual clusters are well sampled to and beyond this magnitude limit. While the results of some of the individual cluster-to-cluster comparisons differ from above, the general picture presented by this test is the same as above.

The results for the NEL GLFs suggest systematic differences among the six clusters. Of 15 independent tests, we would not expect more than one discrepancy at the 2$\sigma$ level if the NEL GLFs were Gaussian-distributed around a common parent. Instead, at least two comparisons differ at the 2$\sigma$ level.

To confirm this, we apply a KS test to the distribution of probabilities from the 15 independent comparisons to examine whether it is normal, assuming Poisson errors in the galaxy counts. This test rules out a normal distribution at $>95\%$ confidence for the NEL GLFs for both mangitude limits, with the distribution being skewed towards lower probabilities than expected from Gaussian statistics. We find no evidence for systematic differences of the overall or EL GLFs among our six clusters.

We have shown the non-universality of the NEL GLF, but have not yet constrained the universality of the GLF in clusters quantitatively. Therefore, we now turn to the second test, the comparison between individual clusters and composite GLFs, to examine to what extent these composites -- overall, EL, and NEL -- can serve as common parent distributions to the six clusters. To ensure that the distribution are statistically independent, we compare each individual cluster to a composite calculated from the five remaining clusters. We calculate the composites as described in \S 3.4. We convolve each composite GLF (in its original form as a bivariate, luminosity-surface brightness distribution) with the Galactic extinction and sampling fraction applicable for the individual cluster. Thus, we predict how many galaxies should have been sampled in that cluster as a function of absolute magnitude if they had been drawn from the assumed parent distribution. This process is essentially the reverse of the SWML algorithm that we use to calculate the composite GLFs. We then project the predicted distribution onto the $M_{R}$ axis and use a KS test and a $\chi^{2}$ test to compare the distributions. In our case, the KS test is more sensitive to systematic discrepancies between the distributions, and therefore we base our discussion on its results. Again, we impose our common absolute magnitude limit of $M_{R}=-18.35$ on all cluster-vs.-composite comparisons; this is the absolute magnitude corresponding to our standard apparent magnitude cutoff of $m_{R}=18$ for the most distant cluster, A3266. 

Most clusters are consistent with the composites formed from the five remaining clusters. The exceptions are the EL and NEL populations of A1060. The KS-test probabilities are $p=0.04$ with 52 galaxies for the NEL GLF, and $p=0.003$ with 10 galaxies for the EL GLF. As this cluster is the lowest redshift cluster in our sample, this result raises the question of whether the fixed angular sampling radius of the fiber spectrograph has introduced an inhomogeneity into the sample by truncating the clusters at different physical radii. 

\subsubsection{Radial Sampling Bias}

To address the possibility of radial sampling bias, we determine new composite GLFs, for which we truncate the higher-redshift clusters to the same fraction of the virial radius sampled by the lower-z clusters. In virialized systems, the virial radius scales with $\sigma$ \citep{girardi}. We therefore scale the angular sampling radii of the low-redshift clusters by $\sigma/D_{A}$, where $D_{A}$ is the angular diameter distance, to find the correct angular truncation radius for the other clusters. The composite GLFs truncated to the A1060 and A496 sampling radii are indistinguishable from the default composite GLFs under a $\chi^{2}$ test. (This is not surprising for two reasons: Not only are the samples dominated by galaxies in the central regions of the clusters, but the scaling of the virial radius with $\sigma$ and the angular scale with $D_{A}$, by coincidence, mostly cancel each other, making the default angular sampling radii of most clusters comparable.) 

We now repeat the cluster-vs.-composite analysis from \S 4.1.1 with these truncated GLFs to find out whether the discrepancies between the composites and the EL and NEL populations in A1060 persist. The A1060 EL and A496 NEL GLFs are now inconsistent with the composites, but the A1060 NEL GLF is consistent. Therefore, we cannot rule out that the small sampling radius is at least partly responsible for the inconsistency in the A1060 NEL GLF that we observed with the untruncated samples. On average, however, we still find five out of six clusters to be consistent within 2$\sigma$ with the composite for the EL and NEL samples, and six out of six for the overall samples. We will discuss below what constraints this places on the universality of the shape of the GLF in clusters.

\subsubsection{Aperture Bias}

Another potential source of inhomogeneity in the sample is aperture bias. Due to the finite angular radius of the $3^{\prime\prime}$ fibers, the spectra only sample light within a limited physical radius around the center of a given galaxy. The spectrum may thus not be representative of the galaxy as a whole, and, in particular, the [OII] equivalent width may be biased low. This effect is obviously of greater concern at lower redshift and for galaxies with large bulges. Therefore, we have to consider whether aperture bias may have enhanced the bright end of the NEL GLF in the most nearby clusters by causing misclassifications of EL galaxies as NEL galaxies.

We have found no strong indications that aperture bias is responsible for the discrepancies observed in the A1060 EL and NEL and A496 NEL GLFs. Although the deviation responsible for the disagreement of A496 with the composite is indeed an excess of galaxies at the bright end and a deficit around $M_{R}\approx-18.5$ (compared to a composite normalized to minimize $\chi^{2}$), the same systematic deviation is qualitatively observed in the A496 overall GLF, which cannot be affected by aperture bias. In addition, the discrepancy between A1060 and the composite is not in the same sense as that between A496 and the truncated composite, even though we would expect a stronger effect for this, more nearby, cluster.

We examine how many galaxies would need to be misclassified in A496 to explain its deviant GLF by aperture bias. Because of the larger physical size of the most luminous galaxies, aperture bias would affect the bright end of the NEL GLF most for a given cluster. The observed deviation would require $\sim40\%$ of the presumed NEL galaxies around $M_{R}\approx-20.5$ to be EL galaxies. Given the small number of identified EL galaxies, this would require $\sim85\%$ of the EL population to have been misclassified as NEL galaxies. For a sample very similar to this, \citet{zaritsky95} estimate that at most $20\%$ of spiral galaxies with $cz\leq15000$ km/s might be affected by aperture bias resulting in a misclassification of their emission line properties. Therefore, judging primarily from the case of A496, aperture bias is unlikely to be the cause of the observed discrepancies between individual clusters and the composites, or between individual cluster NEL GLFs. Aperture bias cannot explain the deviation of the A1060 NEL GLF from the composites, as the composite overpredicts, rather than underpredicts, the number of bright NEL galaxies, and underpredicts the number of faint galaxies.

\subsubsection{The Uniformity of Cluster GLFs}

We have demonstrated in \S 4.1.1 that $\chi^{2}$ comparisons among individual clusters rule out a universal shape of the NEL GLF. The comparisons among individual clusters and the composite GLFs in the previous sections yield additional constraints on the degree of uniformity of GLFs in clusters. 

For our comparisons between individual clusters and the composite GLFs formed from the five remaining clusters, we find all six clusters to be consistent with the composite within $2\sigma$ in the case of the overall GLFs, and five out of six consistent within $2\sigma$ when comparing EL or NEL populations. This allows us to place constraints on the ``universality'' of our composite GLFs, {\it i.e.}, on the fraction of all clusters (with selection criteria similar to those in our sample) that are consistent with our composite GLFs within $2\sigma$. To establish the universality of our composite GLF, that fraction would have to be shown to be at least 0.95 ({\it i.e.}, on average only one cluster out of 20 should show a discrepancy at the 2$\sigma$ level), which is impossible to prove with a sample of just six clusters. However, simple binomial statistics show that it is unlikely ($p<0.05$) to draw five consistent clusters out of a sample of six unless at least $\sim40\%$ of all clusters are in agreement with our composite GLFs. We therefore conclude that our EL and NEL composite GLFs are representative of at least $\sim40\%$ ($2\sigma$ lower bound) of all clusters. Drawing six consistent clusters out of a sample of six, as in the case of the overall GLF, is unlikely unless at least $\sim60\%$ of all clusters are consistent with the composite.

We therefore adopt this fraction of $\sim60\%$ for the overall GLF and $\sim40\%$ for the EL/NEL GLFs as lower bounds on the fraction of all clusters for which our composites are representative. While this argument does not prove the existence of a universal GLF, it indicates a degree of uniformity that is interesting given the relatively wide range of velocity dispersions spanned by these six clusters.  

Larger cluster samples, analyzed in a similar way, would provide tighter bounds on the universality of our composite GLF, provided that each cluster is sampled deeply enough to provide significant constraints on its consistency with the composite. It is unlikely, however, that the limit would be revised downward from $\sim60\%$. The sampling in each of our six clusters is already almost complete over the magnitude range considered here. Furthermore, the uncertainties associated with the composite GLF, which are already much smaller than those associated with the individual clusters, are not propagated through the KS test, so that the lower bound that we derive from this test is a conservative one.

 It is possible that differences among cluster GLFs exist at the very faint end, which has been sampled by us only in the nearest clusters. Given that the comparisons between NEL GLFs yield the strongest indications for non-random discrepancies (stronger than for the better-sampled overall GLFs), searches for systematic differences among clusters are more likely to succeed if NEL galaxies are considered separately.

\subsection{Cluster Composite GLF}

The right hand panels in Fig. 6 show the composite cluster GLFs for the complete (top row), EL (middle row) and NEL (bottom row) samples, based on all six cluster fields. The numerical values of these GLFs are given in Table 3.

\begin{figure}
\plotone{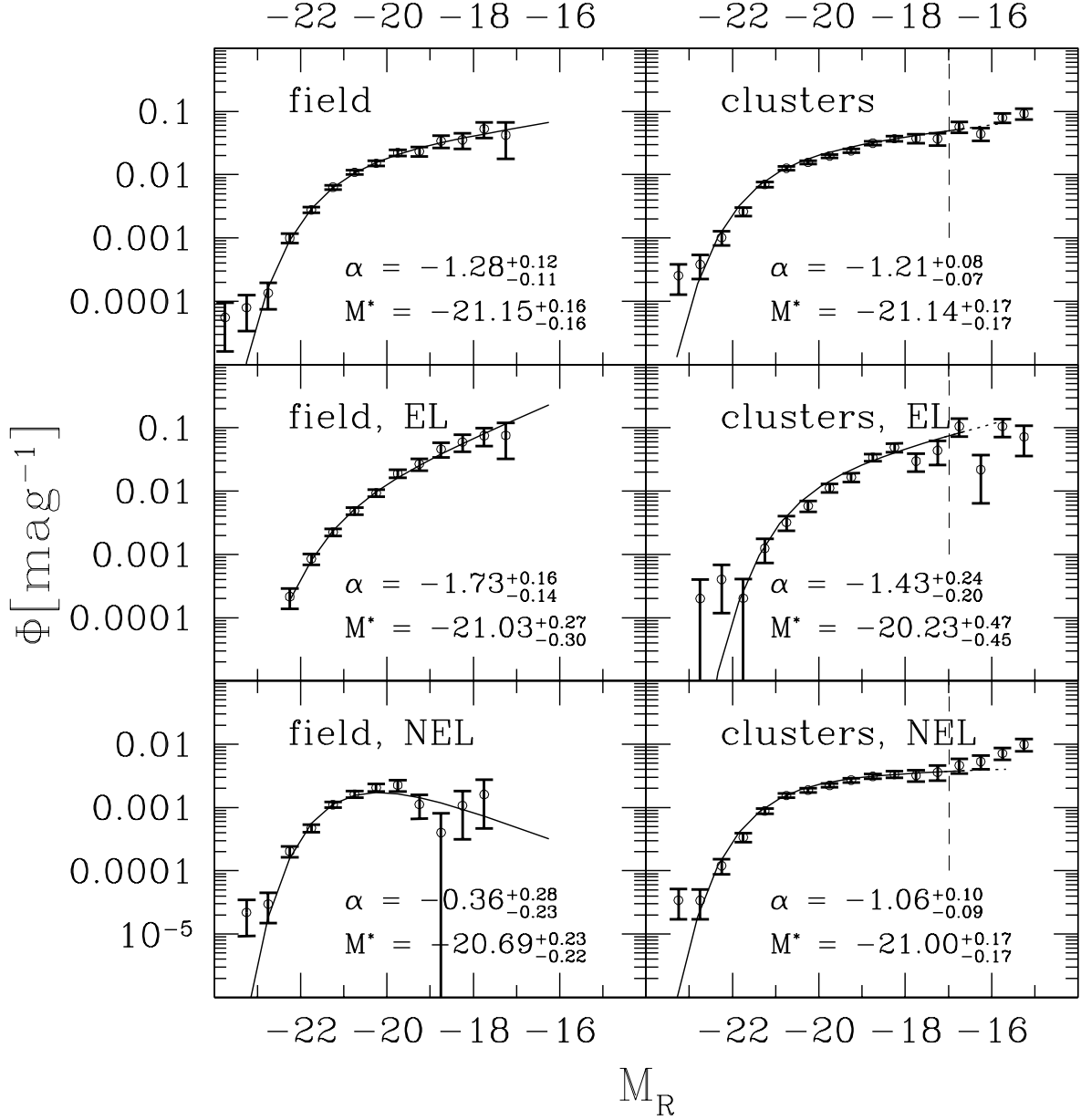}
\caption{Composite GLFs for field and cluster subsamples and different spectral types. Schechter fits down to a magnitude limit of $M_{R}=-17$ mag are also shown. Normalization is arbitrary. Vertical lines indicate limit beyond which only A1060 contributes ($M_{R}\approx-17$ for our adopted apparent magnitude cutoff of $m_{R}=18$).}
\end{figure}

\begin{deluxetable}{lllllll}
\tabletypesize{\scriptsize}
\tablecaption{Composite GLFs. GLF values given in decadic logarithms; number of galaxies given in parentheses.}
\tablewidth{0pt}
\tablehead{
\colhead{$M_{R}$}&\colhead{field, all}&\colhead{field, EL}&\colhead{field, NEL}&\colhead{clusters, all}&\colhead{clusters, EL}&\colhead{clusters, NEL}
}
\startdata
 -23.75& -4.256(2)&  ...&  ...&  ...&  ...&  ...\\
 -23.25& -4.100(3)&  ...& -4.658(3)& -3.595(4)&  ...& -3.467(4)\\
 -22.75& -3.872(5)&  ...& -4.525(4)& -3.418(6)& -3.698(1)& -3.469(4)\\
 -22.25& -3.000(36)& -3.668(8)& -3.692(26)& -2.991(16)& -3.395(2)& -2.920(14)\\
 -21.75& -2.555(91)& -3.071(28)& -3.325(54)& -2.582(41)& -3.693(1)& -2.472(39)\\
 -21.25& -2.198(168)& -2.649(55)& -2.957(100)& -2.154(110)& -2.905(6)& -2.056(100)\\
 -20.75& -1.964(158)& -2.311(62)& -2.789(77)& -1.898(198)& -2.495(15)& -1.811(172)\\
 -20.25& -1.820(120)& -2.030(57)& -2.686(52)& -1.805(238)& -2.234(25)& -1.727(192)\\
 -19.75& -1.653(82)& -1.724(48)& -2.648(24)& -1.704(279)& -1.952(42)& -1.652(195)\\
 -19.25& -1.631(33)& -1.572(22)& -2.950(6)& -1.621(230)& -1.781(41)& -1.568(166)\\
 -18.75& -1.468(20)& -1.338(15)& -3.396(1)& -1.501(243)& -1.468(64)& -1.509(146)\\
 -18.25& -1.449(13)& -1.226(11)& -2.970(2)& -1.431(118)& -1.310(41)& -1.477(66)\\
 -17.75& -1.278(13)& -1.124(10)& -2.795(2)& -1.426(41)& -1.527(10)& -1.492(24)\\
 -17.25& -1.373(3)& -1.120(3)&  ...& -1.433(21)& -1.358(6)& -1.443(14)\\
 -16.75&  ...&  ...&  ...& -1.242(26)& -0.975(10)& -1.335(15)\\
 -16.25&  ...&  ...&  ...& -1.355(19)& -1.663(2)& -1.275(17)\\
 -15.75&  ...&  ...&  ...& -1.100(33)& -0.979(10)& -1.145(23)\\
 -15.25&  ...&  ...&  ...& -1.032(26)& -1.143(4)& -1.005(22)\\
\enddata
\end{deluxetable}

We fit Schechter functions to these GLFs down to a limiting magnitude of $-17$ (the magnitude at which the composite GLF is based on more than one cluster). The best fit Schechter parameters are given in the individual panels of Fig. 6 and are also listed in Table 4. All fits are consistent with the discrete GLFs within 2$\sigma$ over the specified magnitude range.

\clearpage
\begin{deluxetable}{lllllll}
\tabletypesize{\scriptsize}
\tablecaption{Schechter parameters}
\tablewidth{0pt}
\tablehead{ \colhead{Sample} & \colhead{$\alpha$} & \colhead{$M_{R}^{*}$} &\colhead{$p_{chi^{2}}$}}
\startdata
field, all &$-1.28^{+0.12}_{-0.11}$&$-21.15^{+0.16}_{-0.16}$&0.84\\
field, EL  &$-1.73^{+0.16}_{-0.14}$&$-21.03^{+0.27}_{-0.30}$&0.94\\
field, NEL &$-0.36^{+0.28}_{-0.23}$&$-20.69^{+0.23}_{-0.21}$&0.10\\
clusters, all &$-1.21^{+0.08}_{-0.07}$&$-21.14^{+0.17}_{-0.17}$&0.19\\
clusters, EL  &$-1.43^{+0.24}_{-0.20}$&$-20.23^{+0.47}_{-0.45}$&0.10\\
clusters, NEL &$-1.06^{+0.10}_{-0.09}$&$-21.00^{+0.17}_{-0.17}$&0.56\\
\enddata
\end{deluxetable}

Fig. 7 shows the 1- and 2$\sigma$ error contours for the fits, as determined from the method of \citet{avni}, with the $\Delta\chi^{2}$ values taken to be the $\chi^{2}$ values for two degrees of freedom ($\alpha$ and $M^{*}$) and probabilities of 0.315 and 0.05. 

\begin{figure}
\plotone{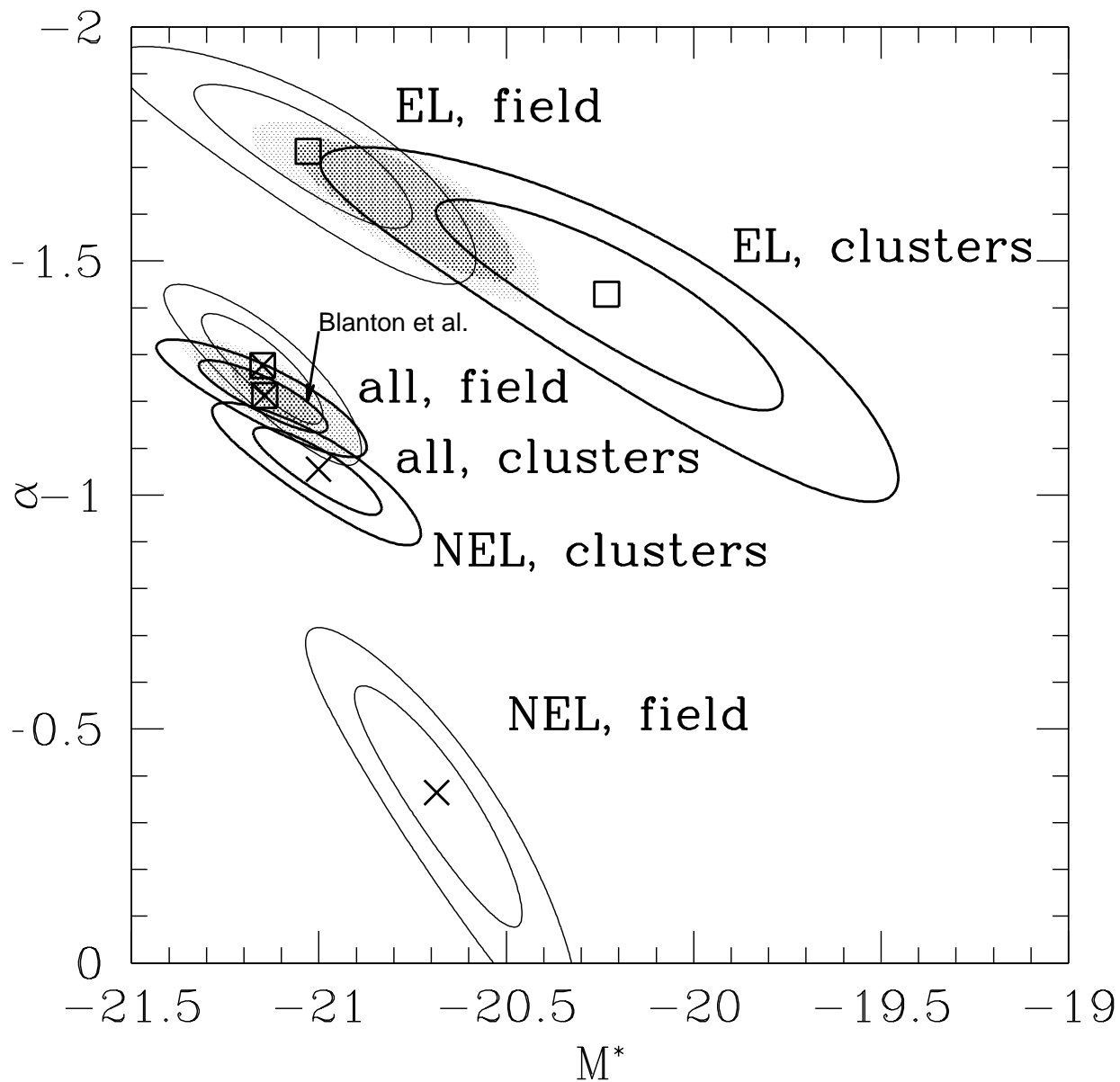}
\caption{$1\sigma$ and $2\sigma$ error contours in Schechter parameter space for our Schechter fits to the combined field and cluster GLFs down to $M_{R}=-17$ mag. The arrow shows the field GLF from \citet{sloanlf} with a transformation $M_{R}-M_{r^{*}}=-0.2$.}
\end{figure}

The failed spectroscopic targets introduce an additional uncertainty in the faint end slope of $\Delta\alpha\approx\pm 0.03$ and $\Delta M^{*}\approx\pm0.01$. The effect of selecting galaxies in a different magnitude band is on the order of $\Delta\alpha\approx+0.02$. The effect of higher-order variations of the sampling fraction is on the order $\Delta\alpha\approx+0.01$. As these corrections are all distinctly smaller than our statistical uncertainties and the sense of the largest of these is unclear, we neglect them.

We have tested whether these composite GLFs represent all of our individual cluster GLFs well by comparing them using the same procedure as in \S 4.1, {\it i.e.}, convolving them with the sampling function and Galactic extinction for any given cluster, and comparing the predicted distribution of sample galaxies to the observed distribution using a KS and $\chi^{2}$ test. As our purpose is only to verify that our composites represent all of the individual cluster GLFs, we carry the comparisons as far as justifiable in each case, {\it i.e.}, to a magnitude limit of $m_{R}=18$ (but not fainter than $M_{R}=-17$, beyond which only A1060 contributes to the composite).

We find that the overall, EL and NEL composites represent the respective galaxy distributions in all six individual clusters well ({\it i.e.}, no comparison --- either by a $\chi^{2}$ or KS test --- shows a discrepancy at the level of 2$\sigma$ or more). Therefore, we conclude that our composite GLFs are good representations of all six clusters in our sample and of $\geq$60\% (as estimated in \S 4.1.4; $\geq$40\% for the EL and NEL GLFs) of all clusters obeying similar selection criteria.

\subsection{Field GLFs}

The sample contains 1527 galaxies not associated with the six clusters, 749 of them within our redshift and magnitude limits. This enables us to calculate GLFs for field galaxies in the same way as our cluster composite GLFs, and to compare them in a self-consistent way.

We apply the SWML method with an apparent magnitude cutoff of $m_{R}=18$ ($m_{R}=17$ for the A3266 field) to calculate the field GLF. We impose redshift limits of 6000 $<\:cz\:<$ 50000 km/s. The lower limit is motivated by the fact that there is a low-z ($cz\approx4000$ km/s) feature in the field of A1631 (apparently associated with NGC4756) that would dominate the faint end of the field GLF if included. The distribution of these galaxies in a plot of [OII] EW versus $M_{R}$ is uncharacteristic of field galaxies in the five other fields and at higher redshifts; the high fraction of galaxies with very low [OII] EW resembles a cluster population rather than a field population. While the field sample is supposed to cover a representative range of environments, including higher-density ones, the volume of space sampled at such low redshifts is not large enough to guarantee a representative sample of the field population if this feature is included. 

The composite field GLFs are presented in the left hand column of Fig. 6 and in Table 4. As for the cluster composite GLFs, we fit Schechter functions for $M_{R}\leq-17$. The 1- and 2$\sigma$ error contours are given in Fig. 7. 

Our results for the field GLF are in agreement with the values given by \citet{sloanlf} for the Digital Sky Survey (SDSS) of $\alpha=-1.20$ and $M_{r^{*}}=-20.83$ (assuming $M_{r^{*}}-M_{R}\approx0.2$, based on \citet{fukugita}). If we adopt a reasonable color of $(B-R)\approx1.0$, our field GLF is also in agreement with the determination of \citet{2dflf} from the 2dFGRS.

As in the case of the cluster GLFs, the various biases afflicting the calculation of the sampling fraction do not affect the composite field GLF significantly. The failed spectroscopic targets introduce worst-case uncertainties of $\pm0.03$ in $\alpha$ and $M^{*}$ in the field GLF. Systematic effects in $\alpha$ from the color selection bias and higher order variations in $m_{R}$ are on the order of $\Delta\alpha\leq0.01$ each.

\subsection{Comparisons between Field and Clusters}

We use three different methods to compare our composite cluster and field GLFs to each other. The first is a simple $\chi^{2}$ comparison of the SWML solutions for the overall, EL and NEL samples. We compare them to a limiting absolute magnitude of $M_{R}=-17$, the magnitude to which the cluster GLF is determined from more than one cluster and to which the field GLFs contain data. Table 5 summarizes the results of these comparisons. The overall and EL GLFs are consistent between the field and clusters, but the field NEL GLF is clearly distinct from the cluster NEL GLF at more than 3$\sigma$.

To test for systematic differences in the shape of the GLF to which the $\chi^{2}$ test may be insensitive, we also compare the composite GLFs in terms of their Schechter fits. Two GLFs are obviously inconsistent if the 2$\sigma$ error contours of their Schechter fits are disjoint. If that is not the case, a more differentiated evaluation is required. If the error countours overlap, any set of parameters within this overlap region would be individually consistent with both GLFs. However, the requirement for the correct Schechter values of both GLFs to lie simultaneously within this overlap region imposes an additional constraint on the likelihood of such a fit; even a point that is marginally consistent with both GLFs individually may thus not necessarily qualify as a likely simultaneous fit to both.

 To avoid the difficulties of having to calculate the exact probability density for both fits in Schechter space, we approximate this probability by calculating the joint $\chi^{2}$ probability of two different GLFs having been drawn from the same Schechter function. The joint probability is simply the product of the two individual probabilities, renormalized so that the best fit probability for each individual fit is 1. We apply this renormalization because, for the purposes of this comparison, we are not interested in the quality of the fits, but merely in using them as tools to characterize the shapes of the GLFs. (This idea is similar to the procedure for determining the error contours, which are always drawn relative to the minimum $\chi^{2}$, regardless of the quality of the best fit itself.) If a point exists in $(\alpha;M^{*}_{R})$ space for which the joint probability is high enough so as not to exclude a simultaneous fit ({\it i.e.}, $>$0.05), and if that point lies within the error contours of both individual Schechter fits, we consider the two GLFs consistent.
Deriving the joint probability by multiplying the two individual probabilities implies that the two realizations of the GLF are independent; therefore, we cannot use this procedure to compare, for example, the NEL composite to the overall composite GLF.

The shaded regions in Fig. 7 indicate regions of the $(\alpha;M^{*})$ plane where the Schechter function is in simultaneous agreement with two different GLFs, and Table 5 lists the results of this test. The parametric comparisons confirm the result from the $\chi^{2}$ test; the field and cluster overall composite GLFs are consistent with each other, as are the EL composite GLFs. {\it However, the NEL GLF differs between the field and clusters under both tests.} Inspection of Fig. 7 reveals that it is steeper in clusters than in the field.

\begin{deluxetable}{lll}
\tabletypesize{\scriptsize}
\tablecaption{Field vs. Cluster comparisons. Probabilities of same parent distributions under $\chi^{2}$ and parametric comparisons.}
\tablewidth{0pt}
\tablehead{
\colhead{Comparison}&\colhead{$\chi^{2}$-test}&\colhead{parametric}
}
\startdata
Field vs. Clusters&0.77&0.83\\
Field EL vs. Clusters EL&0.87&0.31\\
Field NEL vs. Clusters NEL&{\bf $<$0.0005}&{\bf $<$0.0005}\\
\enddata
\end{deluxetable}

To confirm this important conclusion, and maintain consistency with the procedure used earlier to compare individual clusters to the composite GLFs, we apply a third test: can the field GLFs serve as parent to the six individual clusters? The procedure is the same as for the cluster-composite comparisons in \S 4.1, except that we now adopt the field GLFs (EL, NEL and overall) as the hypothetical parent distributions for the six cluster samples. As we did for the comparison between the cluster composite GLF and the individual clusters, we impose a magnitude limit of $M_{R}=-17$ and $m_{R}=18$ (except for the NEL GLF, where we do not have any data for $M_{R}\approx-17.25$ and therefore restrict the comparison to $M_{R}\leq-17.5$). Table 6 shows the results of this test.

\begin{deluxetable}{llll}
\tabletypesize{\scriptsize}
\tablecaption{Comparisons between individual clusters and field GLF, $m_{R}\leq18$}
\tablewidth{0pt}
\tablehead{
\colhead{Cluster}&\colhead{$p_{KS}$}&\colhead{$p_{\chi^{2}}$}&\colhead{\# of galaxies}
}
\startdata
\hline
\multicolumn{4}{c}{overall GLF}\\
A1060 & 0.11  & 0.82 & 101 \\
A496  & 0.08 & 0.51 & 237 \\
A1631 & 0.60  & 0.62 & 272 \\
A754  & 0.81  & 0.94 & 367 \\
A85   & 0.28  & 0.93 & 301 \\
A3266 & 0.41 & 0.53 & 391 \\
\multicolumn{4}{c}{EL GLF}\\
A1060 & 0.06 & 0.71 & 20 \\
A496  & 0.88  & 0.94 & 49 \\
A1631 & 0.20  & 0.79 & 61 \\
A754  & 0.11  & 0.78 & 52 \\
A85   & 0.14  & 0.25 & 43 \\
A3266 & 0.56  & 0.96 & 62 \\
\multicolumn{4}{c}{NEL GLF}\\
A1060 & {\bf 0.000} & {\bf 0.000} & 79 \\
A496  & {\bf 0.000} & {\bf 0.026} & 159 \\
A1631 & {\bf 0.000} & {\bf 0.014} & 210 \\
A754  & {\bf 0.000} & {\bf 0.000} & 310 \\
A85   & {\bf 0.000} & {\bf 0.001} & 201 \\
A3266 & {\bf 0.000} & {\bf 0.047} & 232 \\
\enddata
\end{deluxetable}

The overall field composite GLF is consistent with the cluster populations in all six cases, confirming our earlier finding that the field and cluster GLFs are consistent in shape. The distribution of probabilities is consistent with a normal distribution at $p=0.74$. This conclusion also holds for the EL GLFs ($p=0.1$). However, the field NEL GLF is clearly inconsistent with the NEL populations in all six clusters.

Therefore, our conclusion, supported by all three tests, is that the NEL GLF differs between the field and clusters and is steeper in high-density environments. We cannot detect a systematic discrepancy between the field and cluster EL GLFs (but caution that the constraint is weaker because of the smaller number of EL galaxies).

Before we consider possible implications of this, we have to consider the potential role of aperture bias in introducing an inhomogeneity between the field and cluster NEL samples. The cluster sample, on average, is at lower redshift than the field sample, and is thus more susceptible to aperture bias, which might artificially increase the number of galaxies classified as NEL and raise the faint end slope of the NEL GLF. However, if we limit the field NEL GLF to $cz\leq20000$ km/s, the faint end slope does not steepen. If we impose the same lower redshift limit of $cz\geq6000$ km/s on the cluster NEL GLF as on the field GLF, $\alpha$ does not become significantly shallower. Furthermore, we can perform a simple plausibility check: The difference between the field and cluster NEL GLF within the magnitude range $-20<M_{R}<-18$ amounts to $>60\%$ of the integral of the cluster NEL GLF in this range. This ratio implies that over 350 cluster EL galaxies would have to have been misclassified as NEL galaxies to explain the discrepancy. The error rate in identifying EL galaxies would then have to be $>90\%$. For the reasons given in \S 4.1.3, this large an error is unlikely.

The sense of the variation of the NEL GLF is a steepening of the faint end slope from $\alpha=-0.36$ in the field to $\alpha=-1.06$ in clusters. This observation agrees with \citet{christlein2000}, who also found a steepening of the NEL GLF faint end slope (measured to $M_{R}=-17.5$) in increasingly denser environments. In contrast to \citet{christlein2000}, our overall GLF is not significantly different between low- and high-density environments\footnote{This discrepancy may result from the different survey parameters of the Las Campanas Redshift Survey, particularly its surface brightness cutoff, which has been demonstrated by \citet{sloanlf} to produce an inaccurate field GLF.}. It is interesting that the overall GLF shows no significant difference between the field and clusters, despite the differences between the NEL GLFs and the well established morphology-density relation \citep{dressler80}. In our data, the effects of the morphology-density relation, in which early-type galaxies with their intrinsically shallower GLF are more abundant in denser environments, and the steepening of the faint end slope of the NEL galaxies (which are mostly early-type galaxies) cancel each other within our margins of uncertainty.

Given that the overall GLFs in the field and clusters are similar, the difference between the galaxy populations of these two environments is revealed by the difference between the fraction of EL (or, equally, NEL) galaxies with environment (Fig. 8). To construct this figure, we normalize our GLFs to appropriate units (galaxies per comoving Mpc$^{3}$ for the field; galaxies per cluster for the clusters) and calculate the fraction of EL galaxies as a function of $M_{R}$. We display this fraction both as the ratio of the EL GLF to the sum of the EL and NEL GLF, and as the ratio of the EL GLF to the overall GLF. Because the EL, NEL and overall GLFs are calculated and normalized independently of each other, the overall GLF is not necessarily exactly identical to the sum of the EL and NEL GLF, but the two methods agree well. We calculate the error bars using a simple Monte Carlo algorithm. We take the observed EL fraction in each bin to be the parent distribution. For 1000 trials, we draw samples of galaxies equal to the total number of observed galaxies in that bin and determine the scatter in the distribution of the simulated EL fractions recovered from these trials. We then use this scatter as our error bars in each bin.

The upper panel of Fig. 8 shows the EL fraction $f_{EL}$ in the field and clusters. The solid lines show the results calculated from the parametric fits, circles give the NEL fraction as calculated from the sum of the EL and NEL GLF, and triangles give the ratio of the NEL GLF to the overall GLF. All three methods agree. As we move from the field to the clusters, the fraction of EL galaxies decreases significantly over almost all magnitudes (with the exception of the bright end, where the EL fraction is almost zero in both environments). The most drastic differences in the EL fraction are at fainter absolute magnitudes. Around $M_{R}=-18$, the field population is almost entirely dominated by EL galaxies, while in our cluster sample, the fraction of EL galaxies has dropped to about one third on average\footnote{The absolute numbers, of course, are likely to be dependent on the sampling radius, but our aim is to demonstrate qualitative variations, and with the exception of A1060, the sampling in our clusters is rather homogeneous.}.

\begin{figure}
\plotone{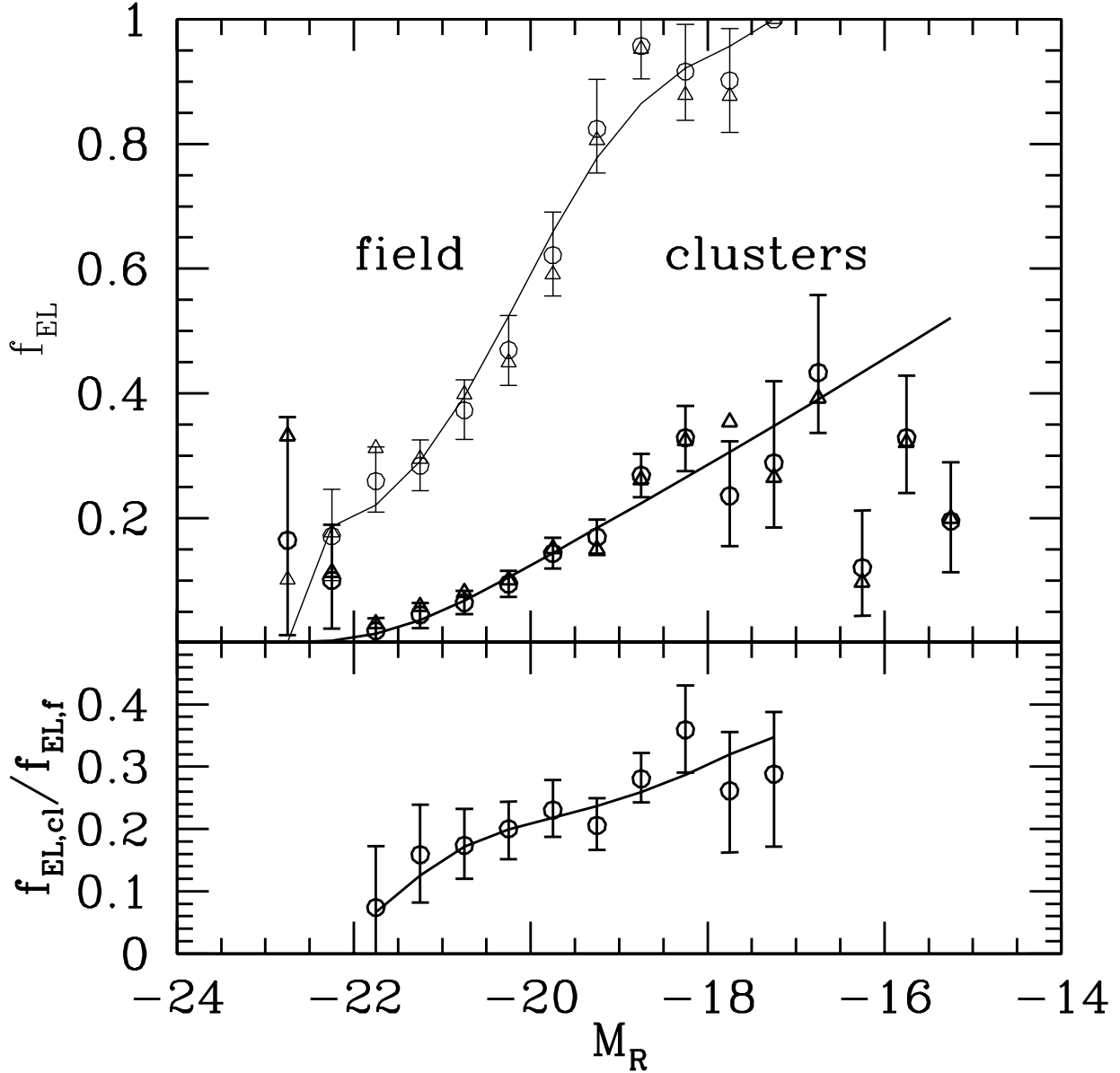}
\caption{Upper Panel: Fraction of EL galaxies as a function of $M_{R}$, derived from the cluster and field composite GLFs. Data points quote fraction relative to sum of EL and NEL GLFs; triangles indicate fraction relative to total GLF. Solid lines are based on Schechter fits. Error bars indicate approximate $1\sigma$ uncertainties, based on Monte Carlo estimate. Bottom panel: Ratio $f_{EL}^{clusters}/f_{EL}^{field}$.}
\end{figure}

 While the change in the EL fraction, $f_{EL}^{field}-f_{EL}^{clusters}$, is larger for fainter magnitudes, $f_{EL}$ itself is also larger at the faint end. It is thus more instructive to plot the ratio $f_{EL}^{cluster}/f_{EL}^{field}$ versus $M_{R}$ to learn whether the properties of dwarf or of giant galaxies are more strongly correlated with their environment. The bottom panel of Fig. 8 shows the ratio $f_{EL}^{cluster}/f_{EL}^{field}$. 

 The curve in Fig. 8 is below one for all magnitudes, indicating that cluster galaxies at all $M_{R}$ are not as likely to be star forming as field galaxies. This is consistent with other recent studies of larger samples \citep{sloansf,2dfsf}. The effect is stronger at the bright end in our data. To confirm this, we have applied a Spearman rank correlation test. The coefficient is $r=0.93$, with a high level of significance ($p<0.0005$ for an accidental correlation). This rules out a zero slope. The best fit slope is positive, indicating that the star formation properties of giant galaxies ($M_{R}\approx M_{R}^{*}$) vary more with the environment than those of fainter galaxies ($M_{R}\geq M_{R}^{*}$).

\citet{depropris}, in their study of the $b_{J}$-band GLF in 60 clusters from the 2dF Galaxy Redshift Survey, obtain results that are complementary to ours. They find a small difference of $\sim0.1$ in the faint end slope $\alpha$ between the field GLF by \citet{2dflf} and their cluster GLF (which is steeper). Such a difference cannot be ruled out from our sample, but could also be due to the greater sensitivity of $b_{J}$ magnitudes to dust and star formation. With $\alpha_{b_{J}}\approx-1.28$, their faint end slope for the cluster GLF is not significantly different from our $\alpha=-1.21$. A small difference in $\alpha$ is expected if there is a color gradient with absolute magnitude ($\alpha_{R}-\alpha_{b_{J}}\approx d(b_{J}-R)/dM_{R}$). If we assign colors of $(B-R)=1.0$ to EL galaxies and $(B-R)=1.5$ to NEL galaxies and use Fig. 8 to estimate the mixing ratio between these two populations at the faint end, the difference in the faint end slope between $b_{J}$-band and R-band GLFs should be $\Delta\alpha\approx0.05$, the blue GLF being steeper than the red GLF. This is entirely consistent with the comparison between these two studies. De Propris et al., comparing their results to the field GLF of \citet{2dflf} also confirm that variations of the GLF in subsamples selected by spectral types are stronger than in the overall GLF, with early spectral types showing a significant steepening of the faint end slope when going from the field to clusters. 

\subsection{The Bright End}

The bright end of the GLF in clusters has been the subject of many studies. Interest has focused on the brightest cluster galaxies, often suggested as a special population \citep{nelson} of standard candles \citep{sandage72,postmanlauer} inconsistent with the Schechter function fit to fainter galaxies \citep{tremaine1977,dressler78}. A related question is whether these galaxies exist in environments other than the densest clusters \citep{morgan}. 

Our cluster overall and NEL GLFs exhibit an apparent excess of galaxies over the Schechter function in the brightest bins ($M_{R}\leq-22.5$). The field GLF also exhibits what may be a bright end excess, but at a lower level. If these galaxies are unique to rich clusters, then there may be a significant difference between the field and cluster bright ends that our previous tests have not been sensitive enough to detect.

To address this question, we normalize the cluster GLF to predict the correct total number of sampled field galaxies with $M_{R}\leq-20$ when integrated over redshift (excluding the redshift ranges associated with the clusters), $M_{R}$, $\mu_{R}$, and the area of the survey. At this bright magnitude limit, any faint end slope differences not affect our normalization. We then use the normalization factor to determine how many field galaxies the cluster GLF would predict in the bright tip with $M_{R}\leq-22.5$. The answer is $\sim22$ galaxies. The observed number of bright field galaxies is 10. The reverse test shows that, when normalized to the number of galaxies in the cluster sample, the field GLF predicts the number of bright galaxies in the clusters to be $\sim4$, whereas the actual number is 10.

We estimate the signficance of this discrepancy in the following way: we assume a certain fraction $\int \phi(M<-22.5) dM / \int \phi(M<-20) dM$ of galaxies in the parent distribution to reside in the bright tip (essentially, we assume a GLF with two bins). Because the number of galaxies with $M_{R}<-22.5$ is negligible compared to the number of galaxies with $M_{R}<-20$, this fraction scales approximately as the fraction of bright galaxies in the spectroscopic sample, $N(M<-22.5)/N(M<-20)$. We then construct 1000 mock samples, each with the same number of galaxies as the actual sample, and use the fraction above (scaled from the bright end fraction of the parent distribution to the bright end fraction of the sample) as the probability that a given galaxy has $M_{R}<-22.5$. From the results, we estimate the probability that the observed number of bright end galaxies or fewer (for the field sample) or the observed number of bright end galaxies or more (for the cluster sample) would have been generated from a luminosity function with this bright end fraction. We then multiply the probabilities for the field and the cluster samples to obtain the probability that the observed bright ends of both samples could have been generated from this luminosity function. We repeat this procedure for a number of bright end fractions.

We find that a common bright end fraction $\int \phi(M<-22.5) dM / \int \phi(M<-20) dM$ for the field and clusters is ruled out at a level of 2$\sigma$. The maximum probability (for the assumption that $\sim1\%$ of galaxies with $M_{R}<-20$ are in the bright tip) is 0.02. Therefore, {\it the brighty end in the cluster GLF is significantly enhanced compared to that of the field}.

Applying the same test to the NEL GLF, we cannot rule out that the bright end fractions in the field and clusters are the same (the probability for drawing the observed numbers of bright galaxies from a GLF with a bright end fraction of $\sim1\%$ is $\sim0.06$). 

\citet{depropris} find $M^{*}_{b_{J}}=-20.07$, but do not explicitly investigate the possibility of an enhancement of the bright end of the GLF. Assuming $(b_{J}-R)\approx1.5$ for early type galaxies, this $M^{*}_{b_{J}}$ is brighter than we would expect from our results. The reason for this discrepancy is not clear.

\section{Conclusions}

 Using a stepwise maximum likelihood algorithm that we have modified to account for sampling fractions that vary both in magnitude and surface brightness, we have calculated R-band galaxy luminosity functions for six nearby clusters, as well as composite cluster and field GLFs, from deep spectroscopic samples. The deepest GLF for an individual cluster, A1060, extends to $M_{R}=-14$ ($M^{*}+7$), making this the deepest spectroscopic survey of the cluster GLF to date. The composite GLF in clusters is consistent with a Schechter function with $M^{*}_{R}=-21.14^{+0.17}_{-0.17}$ and $\alpha=-1.21^{+0.08}_{-0.07}$. Employing the same methods and the same cluster fields, we calculate the composite field GLF using non-cluster members, which allows for a homogeneous comparison of cluster and field environments. The field GLF is best fit with a Schechter function with $M^{*}_{R}=-21.15^{+0.16}_{-0.16}$ and $\alpha=-1.28^{+0.12}_{-0.11}$, in agreement with other recent determinations of the field GLF.

\begin{enumerate}

\item There is a considerable degree of uniformity among the GLFs of our six clusters. We estimate that, at a 2$\sigma$ confidence level, our overall composite cluster GLF is representative for at least 60\% of all clusters obeying similar selection criteria down to $M_{R}=-18.35$. Our composite GLFs for emission line and non-emission line galaxies are representative of at least 40\% of similar clusters.

\item The GLFs of non-emission line (quiescent) galaxies vary significantly among clusters.

\item The overall and emission line (star forming/active) GLFs are indistinguishable between the field and clusters in our sample, except for a significant enhancement in the luminous tip of the cluster GLF ($M_{R}<-22.5$) relative to the field.

\item The GLF of non-emission line (quiescent) galaxies varies significantly between the field and all the clusters, corresponding to a steepening of the faint end slope for $M_{R}<-17$. 

\item The fraction of star forming galaxies varies more strongly with environment for giant galaxies ($\approx M_{R}^{*}$) than for fainter galaxies ($\approx M_{R}^{*}+2$).

\end{enumerate}

\acknowledgments

We thank Dennis Zaritsky for discussions and comments on the manuscript and for providing some of the imaging data, Jose Arenas and Doug Geisler for supplying us with a missing exposure of A496, Bill Tifft for helpful discussions, and an anonymous referee for thoughtful suggestions. AIZ acknowledges support from NAG 5-11108. 
This research has made use of the NASA/IPAC Extragalactic Database (NED) which is operated by the Jet Propulsion Laboratory, California Institute of Technology, under contract with the National Aeronautics and Space Administration. 

The Digitized Sky Survey was produced at the Space Telescope Science Institute under U.S. Government grant NAG W-2166. The images of these surveys are based on photographic data obtained using the Oschin Schmidt Telescope on Palomar Mountain and the UK Schmidt Telescope. The plates were processed into the present compressed digital form with the permission of these institutions. 

The Guide Star Catalog-I was produced at the Space Telescope Science Institute under U.S. Government grant. These data are based on photographic data obtained using the Oschin Schmidt Telescope on Palomar Mountain and the UK Schmidt Telescope.

\clearpage

\appendix

\section {The Detection Catalog}

An accurate survey of galaxy number counts requires a reliable detection algorithm that does not generate spurious detections but still reaches to faint surface brightness limits, good star/galaxy classification, and accurate photometry. For this reason, we decided on an approach combining automatic object detection and extraction with extensive manual control and corrections.

The detection surface brightness threshold is low, particularly on images with higher backgrounds and thus lower S/N, thus generating many spurious detections. In addition, the clusters considered in this paper are at low redshifts, and many of their member galaxies appear as large and highly structured objects. This complicates automatic detection mechanisms such as those provided by Source-Extractor, which is more suited to higher redshift regimes. 
There are two major concerns. The first is the appropriate choice of the background mesh size used by Source Extractor to compute the background count level. Too large a mesh size will not be able to detect small-scale background variations, such as might be caused by scattered light from bright galactic stars. Too small a background mesh, on the other hand, tends to overestimate the background level at the positions of very extended objects. 
The second problem consists of the accidental blending of fainter objects with nearby bright sources -- giant galaxies or bright stars. When run with default parameters, Source Extractor sometimes associates detections with bright objects, even in entirely different parts of the frame. On the other hand, lowering the deblending threshold risks splitting up some of the rather structured galaxy images.

We thus employ a strategy allowing for a maximum of individual control and human intervention. First, we generate four Source Extractor output catalogs. In addition to the default catalog with standard parameters and a background mesh size of 64 pixels, a second catalog uses a larger mesh size of 200 pixels. Photometry for very extended objects (larger than 5000 pixels in isophotal area, consistent with the threshold and mesh sizes used by \citet{tran} and \citet{zabmul00}) is automatically drawn from this second catalog. We generate a third catalog with a low deblending threshold and small mesh size of 32 to detect small scale fluctuations that may be superimposed on -- and consequently confused with -- larger extended objects. A fourth catalog is generated from the cosmic-ray-free frame.

We then cross-correlate the four catalogs, associating detections in each of the four catalogs with their counterparts in the other catalogs, if present. We remove any detection that does not have a counterpart on the cosmic-ray-free image and whose combination of magnitude and surface brightness is characteristic of cosmic rays. We earmark detections that meet only one of these criteria for a visual inspection (such detections are typically spurious -- plate flaws or scattered starlight -- or grazing cosmic rays).

At this point, we identify objects that may have been accidentally blended with brighter sources by the following procedure: We assume that, among the three catalogs, all individual objects in the frame meeting our magnitude and surface brightness limits have been detected. If we find that a detection does not have a direct counterpart in another catalog, but resides in the area occupied by another object in that catalog, the first detection is earmarked as a potential child object, and the other detection as its parent. This could either be a genuine subdetection within a brighter and more extended object, such as an HII region in a large spiral, or it could be a separate object in need of deblending.
To determine these potential parent-child hierarchies, we make use of the segmentation images provided by Source-Extractor. For each of our catalogs, they show which pixels of the input image have been assigned to which object.

We use two automatic methods to classify each object as a star or galaxy. The first of these is the stellarity index provided by Source-Extractor. This is a fractional value between 0 and 1, generated by a neural network algorithm, that is ideally 1 for a star and 0 for a galaxy.
The second mechanism is based on the fact that stars and galaxies occupy two distinct regions in the plot of magnitude versus surface brightness. We visually estimate a linear separatrix between the two regions. We consider objects with lower isophotal surface brightness for a given isophotal magnitude than indicated by this separatrix are galaxies.
Variations in the seeing and other effects on the image quality (such as imperfect tracking) lead us to separately determine parameters for the seeing, the star/galaxy threshold, and the separatrix for each individual frame to optimize the agreement between both methods. 

We generate a catalog of detections that are then subjected to a visual inspection. In each cluster, we perform a full review on a number of frames (several for each night, especially such where star/galaxy separation may be problematic due to bad seeing or tracking), including the central frame. We inspect every detection down to $m_{R}=19.75$ visually (minus the cosmic rays that have been removed automatically). We also perform a limited review on the remaining frames, ignoring the objects that have been classified as stars by both automatic classifiers and reviewing only the earmarked detections plus all galaxies. This approach is justifiable, as the star classifier is usually robust, and the full reviews confirm that almost none of the stellar-appearing objects would be reclassfied as galaxies in a visual inspection.

During the review, we give every inspected object a third, manual classification as a star, galaxy or defect. In addition, we inspect all detected potential child-parent hierarchies and break them up manually if this appears justified. We also have the option of manually associating two separate detections that have not been classified as parent and child.

By comparing the visual appearances of detections of similar magnitudes, we catch blatant systematic errors in the photometry. While the photometry obtained from Source-Extractor is usually reproducible within our quoted errors, galaxies superimposed on bright galaxy envelopes or on scattered light from galactic stars are almost invariably estimated to be too bright. We perform individual photometry on these cases, using a custom-made algorithm that models non-uniform backgrounds over a circular region, and adopt these magnitudes over those from Source-Extractor. In a very small number of cases in which the close proximity of highly structured objects makes a direct determination impossible, we estimate the magnitude visually. Of 25211 reviewed galaxy detections (4251 of which are above our standard analysis threshold of $m_{R}=18$), we estimate 34 magnitudes (none of them at $m_{R}\leq18$). From visual comparisons to galaxies of similar appearance, we estimate the errors on these magnitudes to be $\approx\pm0.25$ mag.
In addition, in cases where stars are so closely superimposed on galaxies that the two have not been detected separately, we manually add the galaxy to the catalog (71 occurrences, 20 of them at $m_{R}\leq18$). We usually estimate magnitudes for these galaxies as described above.
All objects for which we remeasure or estimate magnitudes are marked in our catalog. Altogether, such objects constitute 307 detections, only 10 of them at $m_{R}\leq18$.  

 Although we do not inspect and classify all stars individually, we inspect every detection (including stars) in a collection of thumbnail images in a second pass through the catalog, thus enabling us to account for most misclassifications of galaxies as stars and for star/galaxy blendings as well.

 If the manual corrections to the parent-child hierarchy require an adjustment of the photometry, it is automatically performed by adding or subtracting isophotal fluxes, depending on whether child objects have been deblended or added and whether this affects the magnitude that enters the final catalog. At this stage, we take all other manual corrections into account as well, adding the manual classification as a third star/galaxy classifier to the two automatically generated flags, removing defects, and generating a final catalog.

Despite our careful approach, we may have missed a small number of galaxies, predominantly because they have been blended with closely superimposed or particularly bright galactic stars. An estimate of this effect is provided by cross-correlating the catalog of spectroscopic targets with our detection catalog at a later stage. Typically, we do not detect $<1\%$ of the spectroscopically confirmed galaxies. We trace some of these cases to multiple spectroscopic targettings of the same galaxy, and the rest to the glare of bright foreground stars. We then add these galaxies to the catalog manually, if that has not happened during the normal review process.

The cross-correlation with the spectoscopic catalog also provides a constraint on the star/galaxy misclassifications. We conclude that this effect is less than 1\% of all spectroscopic objects. Most of the misclassifications are associated with star/galaxy superpositions, and only an extremely small number (on the order of one or two objects per cluster) represent real misclassifications -- usually moderately bright, compact galaxies that had been misclassified as stars.

We have tested the effect of uncertainties in the star/galaxy classification on some of our GLFs by alternatively including either all detections not unambiguously identified as stars or all detections clearly identified as galaxies. The GLFs produced by these two extemes are statistically indistiguishable from the standard GLFs. The first scenario steepens the faint end slope of the field GLF by $\Delta\alpha\approx-0.02$ and of the cluster GLF by $\Delta\alpha\approx-0.06$, the second scenario induces changes of $\Delta\alpha\approx+0.02$ and $\Delta\alpha\approx+0.03$, respectively. The reason that is effect is minor --- even under such worst-case scenarios --- is that the sample is dominated by galaxies in regions of the $(m_{R};\mu_{R})$ plane that are unambiguously non-stellar.

\section{Correction of $f_{s}$ for selection in a different color band}

 We selected the spectroscopic targets for our survey using approximate $m_{B}$ magnitudes, while our photometry and analysis use $m_{R}$ magnitudes throughout. While target selection is probably representative of cluster membership for a given $m_{B}$, a color difference $\Delta(B-R)$ between field and cluster galaxies \citep{pimbblet2001} would create a bias in $f_{s}(m_{R};\mu_{R})$. A cluster galaxy of a given $(m_{R};\mu_{R})$ would have a different $m_{B}$, and thus a different probability of having been selected as a spectroscopic target, than a field galaxy. Therefore, our sampling fraction, determined as the average ratio of spectroscopically sampled galaxies to photometric detections in one $(m_{R};\mu_{R})$ bin, would not be representative of the cluster and field populations at $(m_{R};\mu_{R})$.
We quantify this effect by choosing a mean color difference $\Delta(B-R)$ between field and cluster galaxies and assuming that, for a given $(m_{B};\mu_{B})$, the targetting probability $p^{B}$ of a given galaxy is independent of cluster membership. From this assumption, it follows that
\begin{equation}
p_{f}^{R}(m_{R})=p^{B}(m_{R}+(B-R)_{f})=p^{B}(m_{R}-((B-R)_{cl}-(B-R)_{f})+(B-R)_{cl})=p_{cl}^{R}(m_{R}-\Delta(B-R))
\end{equation}
where $p_{f}^{R}$ and $p_{cl}^{R}$ are the probabilities of a field or cluster cluster galaxy being targetted as a function of R magnitude (and surface brightness, which we do not write out here explicitly), and $\Delta(B-R)=(B-R)_{cl}-(B-R)_{f}$.

We now introduce following notation: $p_{ave}^{-}\equiv p_{ave}(m-\Delta(B-R))$ is the mean targetting probability (irrespective of population) at magnitude $m_{R}-\Delta(B-R)$, and $p_{ave}^{0}$ is the corresponding targetting probability at $m_{R}$.

Our aim is to obtain an expression for $p_{f}^{0}$ and $p_{cl}^{0}$, {\it i.e.} the field- and cluster-specific targetting probabilities, assuming a fixed mean color difference between these two populations. If the targetting function can be approximated to first order then the field galaxy targetting probability is, 
\begin{equation}
p_{f}^{0}=1/2 (p_{f}^{+}+p_{f}^{-})
\end{equation}
To fill in the quantities on the right side, we solve the expressions for $p_{ave}^{0}$ and $p_{ave}^{-}$ for $p_{f}^{+}$ and $p_{f}^{-}$, respectively. We start by writing,
\begin{equation}
p_{ave}^{0}=\frac{N_{det,cl}^{0}p_{cl}^{0}+N_{det,f}^{0}p_{f}^{0}}{N_{det}^{0}}=\frac{(N_{det}^{0}-N_{det,f}^{0})p_{f}^{+}+N_{target,f}^{0}}{N_{det}^{0}}=\frac{(N_{det}^{0}-\frac{N_{target,f}^{0}}{p_{f}^{0}})p_{f}^{+}+N_{target,f}^{0}}{N_{det}^{0}}
\end{equation}
In a similar manner, we find 
\begin{equation}
p_{ave}^{-}=\frac{(N_{det}^{-}-\frac{N_{target,f}^{-}}{p_{f}^{-}})p_{f}^{0}+N_{target,f}^{-}}{N_{det}^{-}}
\end{equation}
We solve these two expressions for $p_{f}^{+}$ and $p_{f}^{-}$, respectively, and plug them into (B2) to obtain
\begin{equation}
2p_{f}^{0}=\frac{N_{target,cl}^{0}}{N_{det}^{0}-\frac{N_{target,f}^{0}}{p_{f}^{0}}} + \frac{N_{target,f}^{-}}{N_{det}^{-}-\frac{N_{target,cl}^{-}}{p_{f}^{0}}}
\end{equation}

We solve for $p_{f}^{0}$, using the positive-root solution for the quadratic formula that we obtain in the process.

For $p_{cl}^{0}$, we analogously obtain the equation
\begin{equation}
2p_{cl}^{0}=\frac{N_{target,cl}^{+}}{N_{det}^{+}-\frac{N_{target,f}^{+}}{p_{cl}^{0}}} + \frac{N_{target,f}^{0}}{N_{det}^{0}-\frac{N_{target,cl}^{0}}{p_{cl}^{0}}}
\end{equation}

This expression is undefined if the $(m;\mu)$ plane does not contain detections in regions that are referenced by expressions such as $N_{det}^{-}$ and $N_{det}^{+}$. To avoid this problem, we ignore the gradient of the sampling fraction along the $\mu$ axis and determine $p_{f}^{0}$ and $p_{cl}^{0}$ as functions of $m_{R}$ alone (which is justifiable, as the dependence on $\mu$ is much weaker than on $m$). We then scale all sampling fractions within the corresponding magnitude bin by the value $p_{f}^{0}/p_{ave}^{0}$ (for the field) or $p_{cl}^{0}/p_{ave}^{0}$ (for the clusters) to obtain the corrected sampling fractions for field and cluster galaxies.

As this correction is ultimately dependent on the choice of the average $\Delta(B-R)$ between field and cluster galaxies, we choose not to apply it throughout, but merely use it to estimate the magnitude of the effect. We estimate $\Delta(B-R)$ as a function of $m_{R}$ from a comparison between our R-band photometry and the approximate $b_{J}$-band magnitudes that were used for target selection. For a given $m_{R}$, the average field galaxy is actually slightly redder than the average cluster galaxy, presumably because a field galaxy in our sample is, on average, more distant and thus more massive than a cluster galaxy of the same apparent magnitude. We do not quote the numerical values for $\Delta(B-R)$ here because they are not accurate photometric colors, but they are on the order of a few tenths of a magnitude for brighter galaxies ($m_{R}<16$), and the difference is smaller at the faint end ($m_{R}=18$). This is encouraging, because it indicates that color terms are large only where the sampling is uniformly complete (bright apparent magnitudes), and small where the sampling fraction gradient is strongest (faint $m_{R}$). Therefore, there are no significant biases in the sampling fraction either for bright or faint galaxies. When correcting for the bias, we find a minute change in the slope of the cluster GLF of $\Delta\alpha<0.02$ and of the field GLF of $\Delta\alpha<0.01$, much smaller than the statistical errors. The changes in $M_{R}^{*}$ are of the same order of magnitude.

\section{Higher-order corrections}
By determining the sampling fraction on a discrete grid in bins with finite widths in $(m,\mu)$ space, we neglect potential higher-order variations of the sampling fraction on scales of the bin width. The mean sampling fraction within one bin is only representative of the sampling fraction at the bin center if the densities of photometric detections and spectroscopically sampled galaxies do not vary more steeply than to first order. To estimate the impact that higher-order variations may have on our determination of the sampling fraction, and consequently on our luminosity function, we attempt to model their distribution within the bins to second order in $m$. We restrict ourselves to variations in $m$, which greatly simplifies the calculation and still accounts for the strongest variations of the sampling fraction.

We empirically determine the second-order moments of the galaxies relative to the bin centers and use them to constrain the $0th$ order coefficient of the Taylor-expansion of the detection density. We then use these coefficients, representing the detection densities at the bin center, rather than the total counts of detections across the bin, to calculate the sampling fraction.

We first Taylor-expand the detection density as $\rho=\sum_{n}a_{n}\delta m^{n}$ and define the $m-th$ order moment of the galaxy distribution in one bin as $p_{m}=\int_{-\Delta m/2}^{+\Delta m/2}\rho(\delta m) \delta m^{m} d\delta m$, where $\delta m$ is the magnitude relative to the bin center. The coefficient representing the detection density at the bin center is then given to second order as
\begin{equation}
a_{0}=(c_{22}p_{0}-c_{20}p_{2})/(c_{00}c_{22}-c_{20}c_{02})
\end{equation}
where $c_{mn}$ are the coefficients of the $m-th$ moment of the Taylor expansion, {\it i.e.}, $c_{mn}=(\Delta m/2)^{n+m+1}/(n+m+1)$ if $n+m=even$ and 0 otherwise. 

Empirically, we calculate $p_{m}=\sum_{bin}\delta m^{m}$. 

As in the case of the color selection correction, we do not apply this correction to our calculations (a drawback of using higher-order moments is increased noise), but only use it to estimate the magnitude of this effect. The change in the faint end slope of the field composite GLF is negligible, on the order of $\Delta\alpha\approx0.01$.

\end {document}